\documentclass[aps,twocolumn,showpacs,nofootinbib,footnote,superscriptaddress]{revtex4-2}

\usepackage{custom}
\usepackage{caption}
\graphicspath{{figures/}}

\begin{document}

\title{Nonmonotonic specific entropy on the transition line near the QCD critical point}

\newcommand{\UIC}{Department of Physics, University of Illinois, Chicago, Illinois 60607, USA}
\newcommand{\LQT}{Laboratory for Quantum Theory at the Extremes, University of Illinois, Chicago, Illinois 60607, USA}
\newcommand{\UMD}{Department of Physics, University of Maryland, College Park, Maryland 20742, USA}
\newcommand{\UC}{Kadanoff Center for Theoretical Physics, University of Chicago, Chicago, Illinois 60637, USA}

\author{Maneesha Sushama Pradeep}
\affiliation{\UIC}
\affiliation{\UMD}

\author{Noriyuki Sogabe}
\affiliation{\UIC}
\affiliation{\LQT}

\author{Mikhail Stephanov}
\affiliation{\UIC}
\affiliation{\LQT}
\affiliation{\UC}

\author{Ho-Ung Yee}
\affiliation{\UIC}
\affiliation{\LQT}

\noaffiliation

\begin{abstract}
We investigate the effect of the quantum chromodynamics (QCD) critical point on the isentropic trajectories in the QCD phase diagram. We point out that the universality of the critical equation of state and the third law of thermodynamics {\em require\/} the specific entropy (per baryon) along the coexistence (first-order transition) line to be {\em nonmonotonic\/} at least on one side of that line. Specifically, a maximum must occur. We show how the location of the maximum relative to the QCD critical point depends on the parameters of the critical equation of state commonly used in the literature. We then examine how the isentropic trajectories followed by adiabatically expanding heavy-ion collision fireballs behave near the critical point. We find that a crucial role is played by the sign of the isochoric temperature derivative of pressure at the critical point; this sign determines on which side of the coexistence curve the specific entropy must be nonmonotonic (i.e., has a maximum). We classify different scenarios of the adiabatic expansion that arise depending on the value of the discriminant parameter and the proximity of the trajectory to the critical point.

\end{abstract}
\maketitle

\section{Introduction}

Acquiring knowledge about the phase structure of quantum chromodynamics (QCD) is one of the most important goals of heavy-ion collision experiments \cite{Bzdak:2019pkr}. A prominent feature of the phase diagram is the QCD critical point where the first-order phase transition line separating the hadron resonance gas (HRG) and the quark-gluon plasma (QGP) phases terminates. While various experimental signatures of the QCD critical point have been proposed and are being searched for, commensurate studies regarding the first-order phase transition are considerably less advanced. Therefore, it is essential to establish an understanding of how the expanding QCD matter approaches and undergoes the first-order phase transition.

Ideal hydrodynamics provides the lowest-order approximation of the expanding fireball created in heavy-ion collisions \cite{Bjorken:1982qr}. Due to the scale hierarchy between the size of the fireball and the microscopic QCD scale, the dissipation accompanying the expansion can be considered small and the entropy to be approximately conserved. Since the baryon number is also conserved, the entropy per baryon number $\hat s \equiv s/n$, which we refer to as specific entropy, is approximately constant, even though both densities of entropy, $s$, and of baryon number, $n$, decrease due to expansion. Therefore, given the equation of state (EOS), i.e., the dependence of thermodynamic quantities, such as pressure on temperature $T$ and baryon chemical potential $\mu$, one can identify the trajectories on the QCD phase diagram as lines of constant $\hat s$.

It is well-known that the EOS near a critical point has certain universal properties. Therefore, it is natural to ask what universal properties of the isentropic (constant~$\hat s$)  trajectories follow from the universality of the EOS. The goal of this work is to address this question.

\begin{figure}[t]
\centering
\includegraphics[width=\columnwidth]{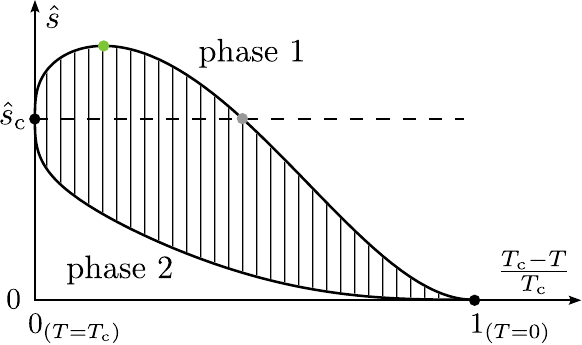}
\caption{A schematic representation of the specific entropy (entropy per baryon) $\hat s$ on the phase coexistence line (first-order phase transition) as a (double-valued) function of the distance from the critical point in terms of reduced temperature. The entropy near the critical point increases on one of the branches (in phase 1) but vanishes at $T=0$. The resulting maximum is denoted by the green dot. The gray dot denotes the point on the branch where $\hat s = \hat s_\crit$ -- the ``critical double.''}
\label{fig:nonmono}
\end{figure}

We start by pointing out that $\hat s$, as a function of the distance from the critical point along the coexistence line, must exhibit a maximum. This is a robust combined effect of the universal behavior of $\hat s$, whose discontinuity on the coexistence line must vanish at the critical point, and the third law of thermodynamics which dictates that~$\hat s$ must vanish at zero temperature. 

The behavior of $\hat s$ satisfying these basic properties is sketched in Fig.~\ref{fig:nonmono}.
The horizontal axis represents the distance from the critical point along the coexistence line in terms of the reduced temperature $(T_\crit-T)/T_\crit$, with~$0$ being the location of the critical point and $1$ -- of the $T=0$ point.  The critical point is a branching point for~$\hat s$, with the leading singular behavior described by
\begin{equation}
\label{eq:sTc-T}
\hat{s} - \hat s_\crit = \pm B_\Gh \left(\frac{T_\crit - T}{T_\crit}\right)^\beta
+\dots\,.
\end{equation}
This universal behavior stems from that of the order parameter field, $\phi \propto \pm \left[ (T_\crit - T )/T_\crit \right]^\beta$, in the conformal theory describing critical phenomena. The exponent $\beta\approx 0.326$ is universal, while the amplitude of the singularity is given by a nonuniversal coefficient $B_\Gh$ depending on the specific details of the theory.
The ellipsis refers to subleading terms.%
\footnote{We shall see that when the maximum is sufficiently close to the critical point these subleading terms are responsible for creating that maximum. The third law is not even needed to establish the nonmonotonicity in this case.} Close to the critical point one of the branches is thus necessarily an {\em increasing\/} function of the distance from the critical point. However, at $T=0$ this function must vanish by the third law. Hence, on this branch, a maximum must occur, as illustrated by the green point on Fig.~\ref{fig:nonmono}.%
\footnote{While Fig.~\ref{fig:nonmono} assumes the simplest scenario where the coexistence line extends all the way to $T=0$, it can be generalized to more sophisticated phase diagram scenarios.}

In order to study this effect quantitatively we consider the range of parameters (relevant for the realistic QCD EOS) in which the behavior of $\hat s$ near the maximum is determined by the competition between the leading term [shown in Eq.~\eqref{eq:sTc-T}] and the subleading singular terms. We employ the critical EOS proposed in Ref.~\cite{Parotto:2018pwx} and reveal the universal properties of the specific entropy, such as the nonmonotonic structure along the first-order boundary (coexistence line). 

We investigate the topography of the specific entropy as a curved surface on a two-dimensional plane of baryon chemical potential $\mu$ and $T$, $\hat s(\mu,\,T)$, and demonstrate the generic presence of a ridge-like structure (see the yellow regions of Figs.~\ref{fig:hat_s_contours} and \ref{fig:hat_s_contours-2}). The nonmonotonic behavior, i.e., the maximum, along the coexistence line illustrated in Fig.~\ref{fig:nonmono} corresponds to the cross section of that ridge.

Such a ridge and the corresponding maximum on the coexistence line emerge either on the HRG side or on the QGP side of the transition.
A simple way to see which side is nonmonotonic is to ask which of the coexisting phases has higher specific entropy at the same temperature (see Fig.~\ref{fig:nonmono}). 
The answer to this question depends on the sign of $B_\Gh$ --- the coefficient of the leading singularity in Eq.~(\ref{eq:sTc-T}). 
We determine this coefficient in terms of the parameters characterizing the mapping of the thermodynamic singularity of the three-dimensional (3D) Ising model to that of the QCD critical point, introduced in Ref.~\cite{Parotto:2018pwx}. 
We also find that the sign determining the nonmonotonic side is the same as that of the temperature derivative of pressure at fixed baryon density [see Eqs.~\eqref{eq:lambda-hill-T} and~\eqref{eq:dPdT} below].

As an example, using the BEST Collaboration EOS with default mapping parameter set choice from Ref.~\cite{Parotto:2018pwx}, we find that the ridge giving rise to the maximum of the specific entropy on the coexistence line emerges on the HRG side. For this scenario, temperature decreases and chemical potential increases as the isentropic trajectory traverses the coexistence region.
Such a scenario is different from the scenario where the system is ``reheated'' on traversing the coexistence region by the entropy released in the process of reconfinement (as in, e.g., Ref.~\cite{Stephanov:1998dy}), but it has been observed in some models of the QCD phase transition without confinement (see, e.g., Ref.~\cite{Scavenius:2000qd}).

The nonmonotonic structure can be characterized by two points: the maximum (green point on Fig.~\ref{fig:nonmono}) and the point where the specific entropy again equals $\hat s_\crit$, which we call ``critical double'' (gray on Fig.~\ref{fig:nonmono}).
Isentropic trajectories can be classified based on the location of the point where they enter the coexistence region (or, if they enter it at all). 
An interesting class is represented by trajectories that enter the coexistence region between the critical point and its double. 
Such trajectories emerge {\em on the same side} of the coexistence line (see Fig.~\ref{fig:class_contours} below).%
\footnote{Such trajectories were seen in the $s$ vs $n$ plane in Ref.~\cite{Akamatsu:2018vjr} (see also Appendix~\ref{sec:n-s_plane}) and possibly on the $(\mu\,,T)$ plane in Refs.~\cite{Nonaka:2004pg,Dore:2022qyz}, although not much attention was given to them.
}

The layout of this paper is as follows. In Sec.~\ref{sec:IsingQCD}, we calculate $\hat s$ near the QCD critical point by using the universal EOS parameterized according to Ref.~\cite{Parotto:2018pwx}. 
In Sec.~\ref{sec:s/n-first} we show how the competition between the leading and subleading terms in Eq.~\eqref{eq:sTc-T} creates a maximum and determines how far it is from the critical point. Section.~\ref{sec:traj} offers demonstrations of $\hat s$ contours on the $(\mu,\,T)$ plane. 
In Sec.~\ref{sec:slope}, we closely examine and classify isentropic trajectories near the first-order boundary. 
We conclude in Sec.~\ref{sec:dis}. 

A series of appendices complements our analysis. Appendix~\ref{sec:susceptibilities} lists the critical susceptibilities used in Secs.~\ref{sec:s/n-first} and \ref{sec:slope}. 
Appendix~\ref{sec:para} reviews the EOS of the 3D Ising universality class.  
Appendix~\ref{sec:s-1st-der} derives specific values of the scaling functions utilizing the findings from Appendix~\ref{sec:para}. 
Appendix~\ref{sec:topo} reviews the mathematical tools used to explore the topography of $\hat s$ on a two-dimensional plane, including the ridge line definition. 
Appendix~\ref{sec:specific-heat} analyses the specific heat singularity and relates it to the singularity in the specific entropy expressed in Eq.~(\ref{eq:sTc-T}). In Appendix \ref{sec:n-s_plane}, we present a geometric description of the maximum specific entropy phenomenon and the classification of isentropic trajectories in the $(n,\,s)$ plane.

\section{Mapping 3D Ising model and QCD}
\label{sec:IsingQCD}
\subsection{The map}
\label{sec:map}

Let us assume the QCD critical point is located at $(\mu_{\rm c},\, T_{\rm c})$ in the temperature vs baryochemical potential plane. 
This point belongs to the 3D Ising universality class, which also includes ubiquitous liquid-gas critical points. Universality means that we can relate thermodynamics near the critical point, i.e., for sufficiently small $(\Delta \mu,\, \Delta T) \equiv (\mu-\mu_{\rm c},\, T-T_{\rm c})$, to the thermodynamics of the Ising model for sufficiently small relevant parameters $(h,\,r)$ (ordering/magnetic field and reduced temperature) related to $(\Delta \mu,\,\Delta T)$ through a linear map \cite{Rehr:1973zz,Nonaka:2004pg,Parotto:2018pwx}:
\begin{subequations}
\label{eq:X-Y}
\begin{align}
\frac{\Delta \mu}{T_{\rm c}} &= - w(r\rho\cos\alpha_1 + h \cos\alpha_2)\,,\\
\frac{\Delta T}{T_{\rm c}} &= w(r\rho\sin\alpha_1 + h \sin\alpha_2)\,,
\end{align}
\end{subequations}
where the notations for dimensionless mapping parameters $w$, $\rho$, $\alpha_1$, and $\alpha_2$ follows Ref.~\cite{Parotto:2018pwx}.%
\footnote{We define parameter $\alpha_2$ as in Ref.~\cite{Pradeep:2019ccv}, which differs from $\alpha_2$ in Ref.~\cite{Parotto:2018pwx} by $180^\circ$.} 
Under this map, close to the critical point, the logarithm of the QCD partition function (i.e., pressure) is equal to the logarithm of the partition function of the Ising model up to and including next-to-leading singularity (i.e., the singularity associated with the energy-like variable $\Gr$.) 

While the universal properties, such as critical scaling exponents, are inherited from the Ising model, the mapping parameters themselves are not universal. 
The parameters $w$ and $\rho$ control the relative scale of the QCD and the Ising variables each measured in terms of their respective critical temperatures.

\begin{figure}[t]
\centering
\includegraphics[bb=0 0 810 960, width=8.5cm]{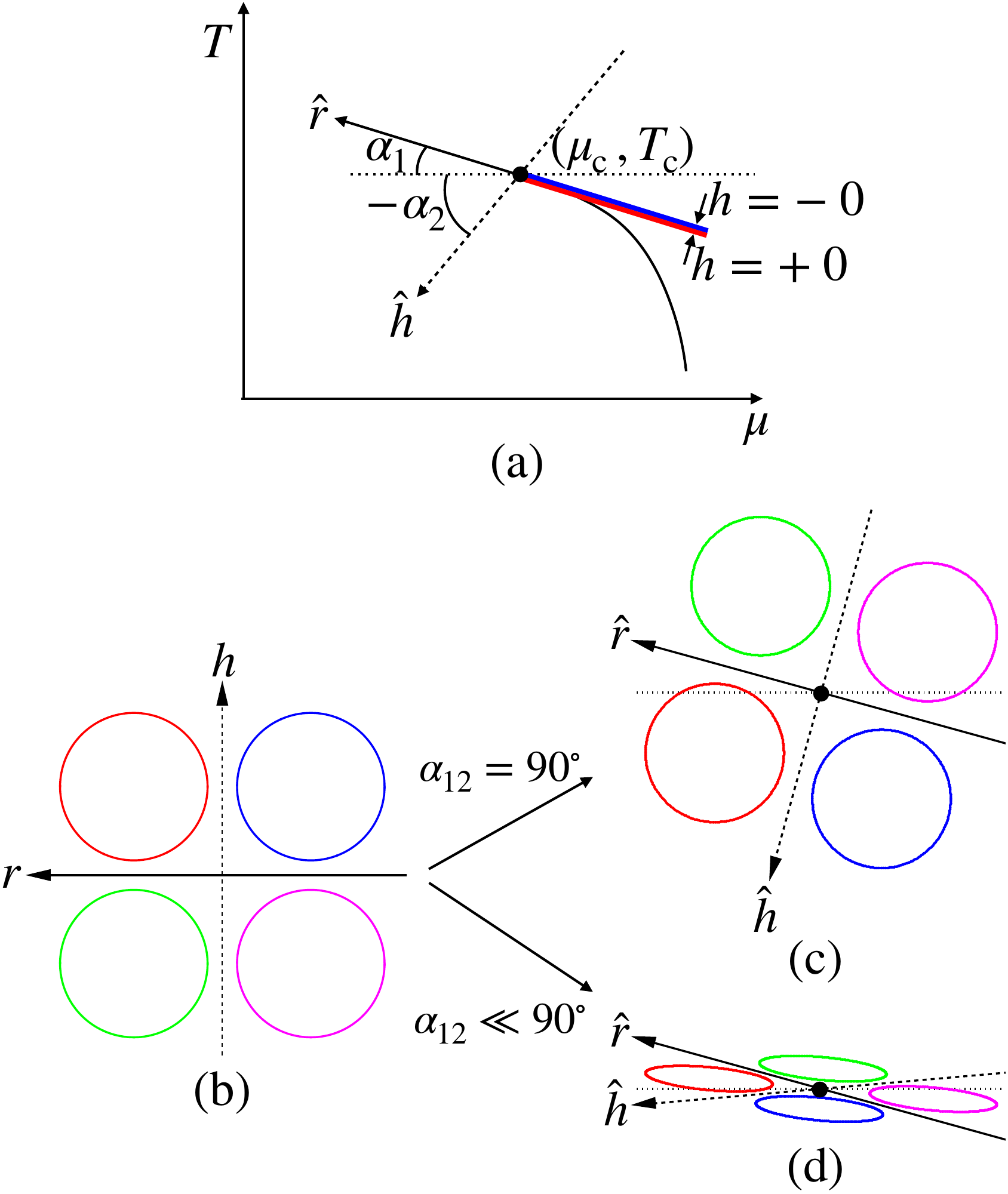}
\caption{Illustration of the role of the angles $\alpha_1$ and $\alpha_2$ in the mapping of the Ising model coordinates $(h,\,r)$ onto the phase diagram of QCD in $(\mu,\,T)$ plane given by Eq.~\eqref{eq:X-Y}. Panels (b)-(d) show the mapping of the points in the four quadrants (color coded) of the $(h,\,r)$ plane onto $(\mu,\,T)$ plane under two different scenarios for the angle difference $\alpha_{12} = \alpha_1-\alpha_2$ used in the literature (see text).
}
\label{fig:map}
\end{figure}

The angle $\alpha_1$ is the (negative of) the slope of the $h=0$ line on the QCD phase diagram at the critical point: 
\begin{equation}\label{eq:tan1}
   \tan\alpha_1=-(\partial T/\partial\mu)_{h=0}.
\end{equation}
This line is the ``$\hat r$-axis" tangent to the coexistence line, as shown in Fig.~\ref{fig:map}. The slope of this line, estimated from lattice calculations, is negative and small. That means $\alpha_1$ is positive and small: $0<\alpha_1\ll90^\circ$. For example, Ref.~\cite{Parotto:2018pwx} finds $\alpha_1\approx 4^\circ$ for their benchmark choice of $\mu_c\approx 350$ MeV.

Similarly, the angle $\alpha_2$ is the negative of the slope of the $r=0$ line (``$\hat h$-axis"): $\tan\alpha_2=-(\partial T/\partial\mu)_{r=0}$. 
Although this line cannot be described geometrically on the phase diagram as easily as the $h=0$ line, the angle $\alpha_2$ or, more importantly, the angular separation $\alpha_{12}\equiv\alpha_1-\alpha_2$ plays an important role in the structure of the EOS near the critical point.

We illustrate the mapping geometrically in Fig.~\ref{fig:map} to compare and contrast two scenarios different by the angle difference $\alpha_{12}$. The default scenario choice of Ref.~\cite{Parotto:2018pwx}, often used in the literature, is $\alpha_{12}=90^\circ$ in Fig.~\ref{fig:map}(c). 
There is no motivation for this choice apart from simplicity. On the other hand, as pointed out in Ref.~\cite{Pradeep:2019ccv}, in the massless quark limit, $m_q\to0$, the angle difference $\alpha_{12}$ vanishes.%
\footnote{The mapping becomes singular (noninvertible) in the $\alpha_{12}\to0$ limit. This reflects the fact that the criticality in this limit corresponds to a tricritical point, in a different universality class \cite{Stephanov:1998dy,Stephanov:2004wx,Pradeep:2019ccv}.} 
Specifically: $\alpha_{12}\to +0$ as $\alpha_{12}\sim m_q^{2/5}$.

This fact, together with the smallness of quark masses, suggests that $\alpha_{12}$ is small and  $\alpha_2<\alpha_1$ in Nature. This is the scenario illustrated in Fig.~\ref{fig:map}(d).

\subsection{Specific entropy}
\label{sec:specific entropy}

We calculate the entropy density $s$ and baryon number density $n$ by differentiating the QCD pressure $P$ expressed, near the critical point, via mapping to the Ising model Gibbs free energy $G$:
\begin{subequations}
\label{eq:s-n}
\begin{align}
\label{eq:P-def-expr}
&P(T,\mu) - P_c = A G(r(T,\mu),h(T,\mu)) + \ldots\,, \\
\label{eq:s-def-expr}
&s-s_c = AG_T +\ldots = A(\Gh h_T + \Gr r_T) + \ldots \,,\\
\label{eq:n-def-expr}
&n-n_c = AG_\mu + \ldots = A(\Gh h_\mu + \Gr r_\mu) +  \ldots \,,
\end{align}
\end{subequations}
with normalization factor $A=T_\crit^4 $ as in Ref.~\cite{Parotto:2018pwx}. The ``$\dots$'' denotes less singular (and regular) terms also vanishing at the critical point. These terms are negligible, sufficiently close to the critical point compared to the singular terms we write out. Specifically, along the coexistence line, $\Gh\sim|r|^\beta$ and $\Gr\sim|r|^{1-\alpha}$, while the omitted terms are at most of order $|r|$.%
\footnote{Given the values of the exponents $\beta\approx0.326$ and $\alpha\approx0.110$, for sufficiently small $|r|$, the hierarchy $|r|^\beta\gg|r|^{1-\alpha}\gg|r|$ holds. We shall focus on such a regime and revisit the justification for neglecting the $\propto |r|$ term in Sec.~\ref{sec:r-linear}.} 
The leading and subleading singularities in Eq.~\eqref{eq:s-n} are due to the two relevant operators of the conformal $\phi^4$ theory corresponding to the magnetization, or order parameter, $\Gh$ and the energy density $\Gr$ \cite{PhysRevE.55.403}:
\begin{align}\label{eq:GhGr}
\Gh \equiv G_h\,, \quad \Gr \equiv G_r\,,
\end{align}
whose $(h,\,r)$ dependence will be discussed further later. Here, the subscripts denote partial derivatives with respect to one of the independent variables in a set, such as $(\mu,\,T)$ or $(h,\,r)$, while the other variable in the set is fixed. We have used the chain rule, $G_X = G_h h_X + G_r r_X$ to obtain the last expression of Eqs.~(\ref{eq:s-def-expr}) and (\ref{eq:n-def-expr}). Partial derivatives of $h$ and $r$ with respect to $\mu$ and $T$ remain constant due to the linearity of the mapping (\ref{eq:X-Y}):
\begin{subequations}
\label{eq:Y_X}
\begin{align}
\left(
\begin{array}{c}
h_\mu \\ 
h_T 
\end{array}
\right)&= - \frac{1}{T_{\rm c} w \sin \alpha_{12}} \left(
\begin{array}{c}
\sin \alpha_1  \\ 
\cos \alpha_1 
\end{array}
\right) \,,\\
\left(
\begin{array}{c}
r_\mu \\ 
r_T 
\end{array}
\right) &= \frac{1}{T_{\rm c} w\rho  \sin \alpha_{12}} \left(
\begin{array}{c}
\sin \alpha_2  \\ 
\cos \alpha_2 
\end{array}
\right) \,.
\end{align}
\end{subequations}
In Eq.~(\ref{eq:s-n}), $P_\crit$, $s_\crit$, and $n_\crit$ represent the constant baseline values at the critical point. They are, of course, nonuniversal and later, in  Sec.~\ref{sec:lattice}, we will be using the values obtained from extrapolating lattice data incorporated in the BEST Collaboration EOS in Ref.~\cite{Parotto:2018pwx}.

It is straightforward to compute the total specific entropy per baryon number, $\hat s = s/n$, using Eqs.~(\ref{eq:s-def-expr}) and (\ref{eq:n-def-expr}). In the regime on which we focus, in the vicinity of the critical point, the leading terms are given by%
\footnote{We shall consider the choice of parameters (see below) which makes the coefficient $\shatGhc$ controllably small. In this regime, the two terms in Eq.~\eqref{eq:shat-approx} are of the same order, $|r|^{1-\alpha}$, while the leading term in ``\dots" is of order $\shatGhc\Gh^2$, which is smaller by a factor of order $\phi\sim|r|^\beta$.}
\begin{align}
\label{eq:shat-approx}
\hat s -  \hat s_\crit = \shatGhc  \Gh + \shatGrc \Gr + \dots\,,
\end{align}
where
\begin{subequations}
\label{eq:hat-s-m_sigma}
\begin{align}
\shatGhc & = \frac{A}{n_\crit}(h_T - \hat s_\crit h_\mu ) = \frac{\hat s_\crit \sin \alpha_1 - \cos \alpha_1}{ w \sin \alpha_{12}}\,\frac{T_{\rm c}^3}{n_\crit}\,,\\
\shatGrc &= \frac{A}{n_\crit}(r_T - \hat s_\crit r_\mu ) =  -\frac{\hat s_\crit \sin \alpha_2 - \cos \alpha_2}{w \rho \sin \alpha_{12}}\,\frac{T_{\rm c}^3}{n_\crit}\,.
\end{align}
\end{subequations}
The coefficients $\shatGhc$ and $\shatGrc$, and in particular their signs, will play a central role in determining the global structure of $\hat s$. 

Since $\sin\alpha_{12}>0$ as discussed at the end of Sec.~\ref{sec:map}, the signs of $\shatGhc$ and $\shatGrc$ are given by
\begin{subequations}
\begin{align}
\label{eq:sign-s-m}
{\rm sgn}\, \shatGhc &=
{\rm sgn}\,(\hat s_\crit-\cot\alpha_1) \cdot\,{\rm sgn}\, (\sin \alpha_1) \,, \\
\label{eq:sign-s-sigma}
{\rm sgn}\, \shatGrc &
=-{\rm sgn}\,(\hat s_\crit-\cot\alpha_2) \cdot\,{\rm sgn}\, (\sin \alpha_2)\,.
\end{align}
\end{subequations}

Since $\alpha_2 < \alpha_1$, the coefficient $\shatGrc$ can be negative only if $\alpha_2$ is in the range, $\arccot \hat s_\crit < \alpha_2 < \alpha_1$. This range does not exist unless $\shatGhc>0$ (i.e., $\hat s_c<\tan\alpha_1$), and even in this case this range is narrow since $\alpha_1$ is small. For this reason, to simplify and focus the discussion, below we shall only consider the scenario with $\shatGrc>0$, i.e.,
\begin{equation}\label{eq:alpha2-range}
    -90^\circ < \alpha_2 < \min(\alpha_1,\arccot \hat s_\crit)\,.
\end{equation}
If needed, our analysis can be generalized to include less likely scenarios with negative $\shatGrc$.

\section{Specific entropy along the coexistence line}
\label{sec:s/n-first}

\subsection{$(r,\,h)$ plane}
\begin{figure}[t]
\centering
\includegraphics[bb=0 0 610 370, width=8cm]{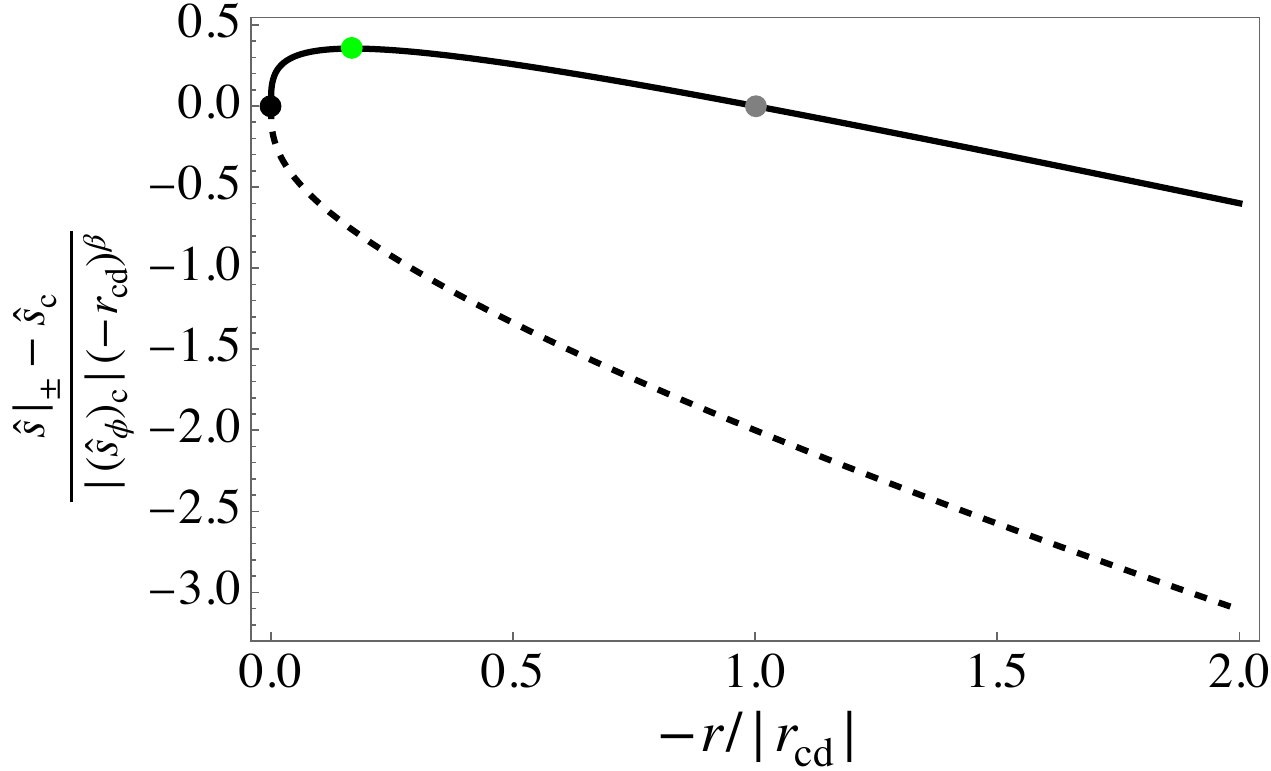}
\caption{Specific entropy along the coexistence line given by Eq.~(\ref{eq:shat-1st}). The black curve represents the branch demonstrating a nonmonotonic behavior, which is one of the main results in the present paper. The dashed curve is the branch that remains monotonic. The black, gray, and green points denote the critical point, critical double, and the maximum of the nonmonotonic branch (see text).}
\label{fig:shat-1st}
\end{figure}
We shall examine $\hat s$ given by Eq.~(\ref{eq:shat-approx}) along the coexistence line (first-order boundary), $r<0$, $h=\pm0$. Close to the critical point, $\Gh$ and $\Gr$ can be expressed in the well-known scaling form (see, e.g., Ref.~\cite{huang2008statistical}):
\begin{align}
\label{eq:m-sigma-sca}
\Gh = (-r)^\beta \tilde \Gh (z)\,, \quad \Gr =  (-r)^{1-\alpha} \tilde \Gr (z)\,,
\end{align}
with the scale-invariant variable $z\equiv h(-r)^{-\beta \delta}$,
where $\alpha \simeq 0.110$, $\beta \simeq 0.326$, and $\delta \simeq 4.80$ are the critical exponents of the 3D Ising model ($\beta\delta = 2-\alpha-\beta$). We shall adopt the same normalization as in Ref.~\cite{Parotto:2018pwx}, which fixes $\tilde \Gh(\pm0) = \pm1$, while $\tilde\Gr(0)$ can be calculated numerically:
\begin{equation}\label{eq:Gr(0)}
    \tilde \Gr(0) \equiv
\tilde\Gr_-(0)\approx -0.66\,, 
\end{equation}
(see Appendix~\ref{sec:s-1st-der}).
Throughout this work, a tilde denotes a scaling function of scale-invariant variable $z$. For the scope of this section, the explicit forms of $\tilde \Gh(z)$ and $\tilde \Gr(z)$ are not important; we need only certain values (in particular, their signs), such as the one given in Eq.~\eqref{eq:Gr(0)}.

The sign of $\tilde\Gr(0)$ can be understood if we calculate the susceptibility, or specific heat, $\Gr_r=G_{rr}$, which must be positive for thermodynamic stability.
At $h=0$:
\begin{equation}
\cx{\Gr_r} = -(1-\alpha)\tilde\Gr(0)(-r)^{-\alpha}>0\,.
\end{equation}
Thus $\tilde\Gr(0)<0$.

Substituting Eq.~\eqref{eq:m-sigma-sca} into Eq.~\eqref{eq:shat-approx} we find:
\begin{align}
\label{eq:shat-1st}
\cx{\hat {s}} - \hat s_\crit = \pm \shatGhc (-r)^\beta + \shatGrc \tilde \Gr(0) (-r)^{1-\alpha} +\ldots\,.
\end{align}
Here, $\cx{}$ denotes evaluation at $h=\pm0$, i.e., on the two sides of the coexistence boundary, at given $r$. 
If the second term is decreasing away from the critical point then on the branch where the first term is increasing there will be a maximum, as shown in Fig.~\ref{fig:shat-1st}. We shall focus on this case, since, as we discuss above, under Eq.~\eqref{eq:sign-s-m}, it corresponds to the more likely scenario, $\shatGrc>0$.

The value of $r$ at which the maximum is achieved is given by (the green point in Fig.~\ref{fig:shat-1st})
\begin{equation}\label{eq:rmax}
    (-r_{\rm max})^{1-\alpha-\beta}
    =\frac{|\shatGhc|}{|\shatGrc|} \frac{\beta}{\Grrtil(0)}\,,
\end{equation}
where $\Grrtil(0)=-(1-\alpha)\tilde\Gr(0)>0$ is the amplitude of the specific heat singularity at the critical point (see Appendix~\ref{sec:susceptibilities}).

While $r_{\rm max}$ depends on the magnitude of $\shatGhc$, the sign of $\shatGhc$ determines the side of the coexistence line (high-$T$ or low-$T$) where the maximum appears, as we discuss in the next section. 
We can determine characteristics of the nonmonotonic behavior, such as the maximum value of the specific entropy at $r=r_{\rm max}$,  
\begin{align}
\label{eq:shat-top}
\hat s_{\rm max} = \hat s_\crit + |\shatGhc| \frac{1-\alpha-\beta}{1-\alpha} (-r_{\rm max})^\beta\,, 
\end{align}
and the value of $r$ at the critical double (the gray point in Fig.~\ref{fig:shat-1st}) where $\hat s = \hat s_\crit $,
\begin{align}
\label{eq:rcd}
(-r_{\rm cd})^{1-\alpha-\beta} = \frac{1-\alpha}{\beta} (-r_{\rm max})^{1-\alpha-\beta}\,, 
\end{align}
which has been used for normalizing the horizontal axis in Fig.~\ref{fig:shat-1st}.

There is a special choice of EOS parameters  such that
\begin{align}
\label{eq:special-para}
\hat s_\crit - \cot \alpha_1\to0\,, 
\end{align}
where the coefficient $\shatGhc$ vanishes (and changes sign), according to Eq.~\eqref{eq:hat-s-m_sigma}. 
At this point $r_{\rm max}$ and $r_{\rm cd}$ vanish. 
This is the regime (sufficiently small $|r|$) where our {\em quantitative} analysis based on the leading critical behavior of the EOS applies. 
It should be kept in mind, however, that even outside of this regime, i.e., when ($|r_{\rm max}|$ is not small), the {\em existence\/} of the maximum is guaranteed, as discussed in the Introduction, by the third law of thermodynamics.

The vanishing of the leading discontinuity of $\hat s$ [$\Delta\hat s\sim\Gh\sim(-r)^\beta$] in the limit given by Eq.~\eqref{eq:special-para} is easy to understand. 
In this limit, the ratio of discontinuities, $\Delta s/\Delta n$ equals the ratio $s/n$, so $s/n$ is continuous. 
This follows from Clausius-Clapeyron law, $\Delta s/\Delta n = -(\partial\mu/\partial T)_{h=0}=\cot\alpha_1$, according to Eq.~\eqref{eq:tan1}.

While angle $\alpha_1$ controls the magnitude of $r_{\rm max}$ in Eq.~\eqref{eq:rmax} via $\shatGhc$, angle $\alpha_2$ also affects $r_{\rm max}$, via $\shatGrc$. Using Eq.~\eqref{eq:hat-s-m_sigma} we find
\begin{equation}
\label{eq:rmax-alpha}
    (-r_{\rm max})^{1-\alpha-\beta}
    = \rho \left|
    \frac{\sin(\arccot\hat s_\crit- \alpha_1)}{\sin(\arccot\hat s_\crit- \alpha_2)}    \right|
    \frac{\beta}{\Grrtil(0)}\,.
\end{equation}
In the range of $\alpha_2$ we focus on, given in Eq.~\eqref{eq:alpha2-range}, $|r_{\rm max}|$ is minimized for $\alpha_2=-90^\circ+\arccot \hat s_\crit$, and it is larger for $\alpha_2 \rightarrow \min(\alpha_1,\arccot \hat s_\crit)$.

\subsection{$(\mu,\,T)$ plane}
\label{sec:mu-T}

We shall now translate the results of the previous section into the $(\mu,\,T)$ plane using the mapping Eq.~\eqref{eq:X-Y}.
On the coexistence line, $h=0$, Eq.~\eqref{eq:X-Y} reduces to
\begin{align}
\label{eq:mu-T-r}
\frac{\Delta \mu}{\Delta \mu_1} =  \frac{-\Delta T}{\Delta T_1} = -r\,,
\end{align}
where, as before, $\Delta\mu\equiv\mu-\mu_\crit$ and $\Delta T\equiv T-T_\crit$, while $\Delta \mu_1$ and $\Delta T_1$ measure the distance of the point, $-r=1$, from the critical point along the $\mu$ and $T$ axes, respectively:
\begin{align}\label{eq:Dmu1-DT1}
\left(
\begin{array}{c}
\Delta \mu_1 \\
\Delta T_1
\end{array}
\right) \equiv
T_{\rm c} w\rho \left(
\begin{array}{c}
\cos \alpha_1 \\
\sin \alpha_1
\end{array}
\right)\,.
\end{align}

Replacing $r$ in Eq.~\eqref{eq:shat-1st} with $\Delta T$ using Eq.~\eqref{eq:mu-T-r} we obtain
\begin{align}
\label{eq:shat-1st-T}
\cx{\hat {s}} - \hat s_\crit = \pm B_\Gh \left( \frac{-\Delta T}{T_\crit} \right)^\beta 
+ B_\Gr \left( \frac{-\Delta T}{T_\crit} \right)^{1-\alpha} 
+\ldots\,,
\end{align}
which is Eq.~\eqref{eq:sTc-T} with subleading singularity written out explicitly.
We now find explicit expressions for the coefficients $B_\Gh$ and $B_\Gr$:
\begin{subequations}
\label{eq:Bs}
\begin{align}
\label{eq:BGh}
B_\Gh & = \shatGhc \left(\frac{\Delta T_1}{T_\crit}\right)^{-\beta} \notag\\
&= \frac{\hat s_\crit \sin \alpha_1-\cos \alpha_1}{w \sin\alpha_{12}} \left(\frac{\Delta T_1}{T_\crit}\right)^{-\beta}
\frac{T_\crit^3}{n_\crit} \,, \\
\label{eq:BGr}
B_\Gr & = \shatGrc \tilde\Gr(0)\left(\frac{\Delta T_1}{T_\crit}\right)^{\alpha-1} 
\notag\\
&= - \frac{\hat s_\crit \sin \alpha_2 - \cos \alpha_2}{w \rho \sin\alpha_{12}}
\left(\frac{\Delta T_1}{T_\crit}\right)^{\alpha-1}
\tilde \Gr(0) \frac{T_\crit^3}{n_\crit} \,.
\end{align} 
\end{subequations}

The maximum of $\hat s$ on the coexistence line occurs at temperature $T_{\rm max}$ given by
[c.f., Eq.~\eqref{eq:rmax}] 
\begin{align}\label{eq:DTmax}
\left(\frac{-\Delta T_{\rm max}}{T_\crit} \right)^{1-\alpha-\beta} = \frac{\beta}{1-\alpha}
\frac{|B_\Gh|}{|B_\Gr|}\,,
\end{align}
where $\Delta T_{\rm max}=T_{\rm max} - T_\crit=r_{\rm max}\Delta T_1$.

The critical double is at temperature $T_{\rm cd}$ given by
\begin{equation}
  \label{eq:DTcd-DTmax}
    (-\Delta T_{\rm cd})^{1-\alpha-\beta} = \frac{1-\alpha}{\beta} (-\Delta T_{\rm max})^{1-\alpha-\beta} \,,
\end{equation}
where $\Delta T_{\rm cd}\equiv T - T_{\rm cd}$.

While we have written down the equations in this subsection in terms of $\Delta T$, one can also express the location of the maximum and the critical double in terms of the distance from the critical point along the $\mu$ direction,  
\begin{equation}
    \Delta\mu = -\Delta T \cot\alpha_1.
\end{equation}

\subsection{The nonmonotonic branch}

As we discussed before, the branch exhibiting the maximum is the one where $\hat s$ is {\em increasing\/} away from the critical point (see Fig.~\ref{fig:nonmono}). To determine whether it is the high-$T$ (QGP) or the low-$T$ (HRG) branch, we turn back to Eq.~\eqref{eq:shat-approx}. The value of $\hat s$ is increasing on the branch where $\shatGhc\Gh>0$. With the choice of the direction of the $h$ axis shown in Fig.~\ref{fig:map}, the HRG branch corresponds to $h>0$.
Thus, given $\Gh_h=G_{hh}>0$ (thermodynamic stability), $\phi>0$ on this branch. Hence, HRG branch ($h\to+0$) is nonmonotonic when $\shatGhc>0$. Using Eq.~\eqref{eq:sign-s-m}  we conclude:
\begin{align}
\label{eq:lambda-hill-T}
{\rm sgn}\, (\hat s_\crit - \cot \alpha_1) = \left\{
\begin{array}{cc}
+1  & 
(\mbox{low-$T$, HRG side}) \\
-1 & 
(\mbox{high-$T$, QGP side})
\end{array}
\right. \,.
\end{align}
Note that, while the binary choice of the direction of the $h$ axis (i.e., the sign of $\sin\alpha_{12}$) in Fig. \ref{fig:map} is arbitrary, the product $\shatGhc\Gh$, and with it Eq.~\eqref{eq:lambda-hill-T}, is independent of that choice.

The quantity determining the nonmonotonic side of the coexistence line in the $(\mu,\,T)$ plane can be expressed in terms of the isochoric temperature derivative of pressure at the critical point,
\begin{align}
\label{eq:dPdT}
\hat s_\crit - \cot \alpha_1 = \frac{1}{n_\crit}
\left(\frac{\partial P}{\partial T} \right)_{\!n,\:\! \crit} \,.
\end{align}
This can be seen as a consequence of the fact that the critical isochore ($n=n_\crit $ line for $T>T_\crit$ is tangential to the coexistence line ($h=0$ for $r<0$) at the critical point, i.e.,%
\footnote{Indeed, the dominant term in $n-n_\crit$ is proportional to~$\phi$, according to Eq.~\eqref{eq:s-n}. Thus the line of constant $n-n_\crit=0$ is the same as the line of constant $\phi=0$, or $h=0$, at the critical point.} 
\begin{align}
\label{eq:sl-n-sl-h}
\left( \frac{\partial T}{\partial \mu} \right)_{n\commacrit} = \left( \frac{\partial T}{\partial \mu} \right)_{h\commacrit} \,.
\end{align}
Using Eq.~\eqref{eq:sl-n-sl-h} with ${\rm d} P = s {\rm d} T +  n {\rm d} \mu$ and Eq.~\eqref{eq:tan1}, one obtains Eq.~\eqref{eq:dPdT}.

We note that for typical liquid-gas critical points, such as in water, $\left. (\partial P/ \partial T\right)_{n\commacrit} >0$, which can be seen directly from the slope of the phase diagram in the ($T,P$) plane.

\subsection{Numerical values from BEST EOS}
\label{sec:lattice}

In this subsection, we shall estimate numerical values for the key quantities we introduced and discussed in the previous sections, such as $r_{\rm max}$, $\hat s_{\rm max}$, etc. Since the QCD EOS is still unknown, we shall use the parametric family of EOS introduced in Ref.~\cite{Parotto:2018pwx}. To facilitate comparison with literature, we shall pick the benchmark parameter choice of Ref.~\cite{Parotto:2018pwx}
and will refer to this set as ``Set 1.'' We shall also consider ``Set 2,'' in which the angle $\alpha_2$ is small, since it is physically motivated in Ref.~\cite{Pradeep:2019ccv}, as we discussed at the end of Sec.~\ref{sec:map}. Each Set 1 and 2 is listed in Table \ref{table:paras}.

\begin{table}[t]
\centering
\caption{Two sets of EOS parameters that we use to illustrate our findings. Set 1 is used in 
Ref.~\cite{Parotto:2018pwx}.}
\begin{tabular}{c|c|c|c|c|c|c}
Set & $\mu_\crit$ (MeV) & $T_\crit$ (MeV) & ~$w$~ & ~$\rho$~ & $\alpha_1$ (${\rm deg}$) & $\alpha_2$ (${\rm deg}$) \\
\hline
1 & 350 & 143.2 & 1 & 2 & 3.85 & $-86.15$ \\
2 & '' & '' & '' & '' & '' & $-5$  \\
\end{tabular}
\label{table:paras}
\end{table}

A crucial role is played by the value of the specific entropy at the critical point, $\hat s_\crit$, which is determined from the extrapolation of lattice data from $\mu=0$. Specifically, for  Sets 1 and 2, one finds
\begin{align}
\label{eq:shat-set}
\hat s_\crit \approx 20.3\,,  19.3\,,
\end{align}
respectively. 

Substituting these values, we find the characteristic parameter values shown in Table \ref{table:estimation}. Here, $r_{\rm max}$ is given by Eq.~\eqref{eq:rmax}, $\hat s_{\rm max}$ -- by Eq.~\eqref{eq:shat-top}, and $\Delta\mu_1$, $\Delta T_1$, translating $r$ into $\mu$ and $T$, -- by Eqs.~\eqref{eq:mu-T-r} and \eqref{eq:Dmu1-DT1}.
\begin{table}[t]
\centering
\caption{Parameters characterizing the maximum of~$\hat s$ on the coexistence line for parameter Sets 1 and 2 from Table~\ref{table:paras}.}
\begin{tabular}{c|c|c|c|c}
Set & $\Delta\mu_1$ (MeV) & $\Delta T_1$ (MeV) & $|r_{\rm max}|$ &
$\hat{s}_{\rm max}-\hat{s}_\crit$ \\ 
\hline
1 & 286 & 19.2 & $9.5\times 10^{-4}$ &
0.076 \\
2 & '' & '' & $2.5 \times10^{-2}$ &
0.89 \\ 
\end{tabular}
\label{table:estimation}
\end{table}

The value of $|r_{\rm max}|\ll1$ in both sets, which means the maximum is close to the critical point, i.e., the approximation assuming that $|r_{\rm max}|$ is small is reasonable. 
The smallness is driven by the smallness of the numerator in Eq.~\eqref{eq:rmax-alpha}, since both $\arccot\hat s_\crit$ and $\alpha_1$ are small. 

The maximum is much closer to the critical point for Set 1. In fact, $\alpha_2$ in this set has a value close to the minimum of $|r_{\rm max}|$ as a function of $\alpha_2$ shown in Eq.~(\ref{eq:rmax-alpha}) (achieved at $ \arccot\hat s_\crit -90^\circ \approx -90^\circ$ for $\hat s_\crit \gg 1$).

On the other hand, the distance from the critical point is significantly larger for Set 2, though it is still small. Here, 
the denominator in Eq.~(\ref{eq:rmax-alpha}) is small as well as the numerator.

The value of $\cot\alpha_1\approx14.9$ means that the value of $\hat s_\crit-\cot\alpha_1\approx 5.$ is positive. This parameter determines the side of the coexistence line where the maximum occurs. According to Eq.~\eqref{eq:lambda-hill-T}, the maximum occurs on the HRG side.

To determine how robust this conclusion is, in Fig.~\ref{fig:shat-best}, we extend the comparison for each Set 1 and 2 to a range of $\mu_{\rm c}$. It can be observed that $\hat{s}_\crit - \cot \alpha_1 >0$ across the entire range of $\mu_\crit$.

\subsection{Justification for neglecting the regular terms}
\label{sec:r-linear}

\begin{figure}[t]
\centering
\includegraphics[bb=0 0 580 370, width=8cm]{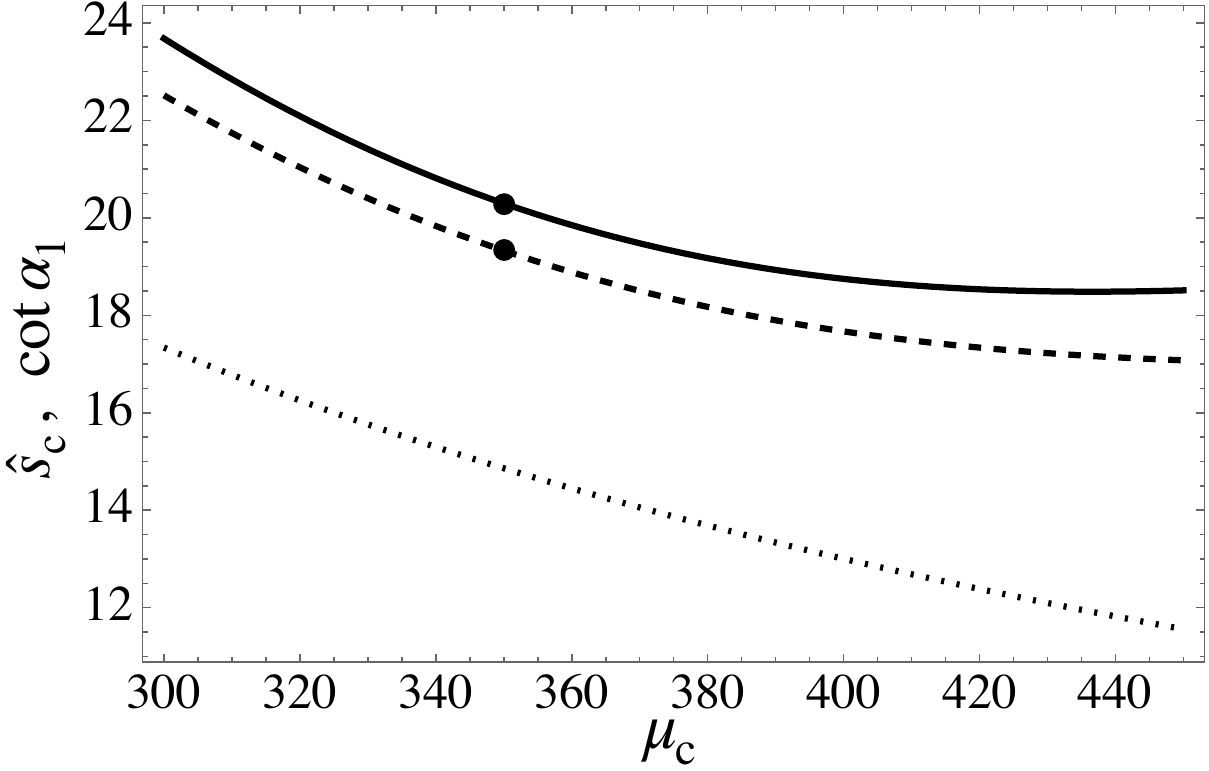}
\caption{Critical specific entropy $\hat s_\crit$ in the parametric family of EOS from Ref.~\cite{Parotto:2018pwx} as a function of the critical parameter $\mu_\crit$ while $w$, $\rho$, and $\alpha_2$ parameters are fixed as in Table \ref{table:paras} for Set 1 (solid line) and Set 2 (dashed line). Black circles represent values in Eq.~\eqref{eq:shat-set}. The dotted line represents $\cot \alpha_1$ as a function of $\mu_\crit$. Since $\hat s_\crit > \cot\alpha_1$ these values support the scenario where the nonmonotonic side appears on the HRG side, according to Eq.~(\ref{eq:lambda-hill-T}). 
}
\label{fig:shat-best}
\end{figure}

Let us check the validity of our assumption that the regular terms $\hat{s}_{\rm reg}$, i.e., the ellipsis in Eq.~\eqref{eq:shat-approx}, are small.
For small $|r|$, its leading term is   
\begin{equation}\label{eq:sreg}
   \hat{s}_{\rm reg} \simeq - \left( \hat{s}_{\rm reg} \right)_{r,\crit} (-r) + \mathcal O(|r|^2)\,. 
\end{equation}
 The Taylor expansion of the pressure around $\mu=0$ used in Ref.~\cite{Parotto:2018pwx} is given by $P_{\rm reg}/T^4  = c_0 (T) + c_2 \left({\mu}/{T}\right)^2 +c_4(T) \left({\mu }/{T}\right)^4 $. 
Here, the coefficients $c_i(T)$ ($i = 0, 2, 4$) are determined from the lattice results (see Ref.~\cite{Parotto:2018pwx} for details).%
\footnote{More precisely, in Ref.~\cite{Parotto:2018pwx}, the regular contribution is defined by subtracting the Ising contribution from the lattice Taylor series. The net contribution is referred to as the non-Ising contribution. Implementing this detail is straightforward and does not impact the order of magnitude estimation in Sec.~\ref{sec:r-linear}. The $\alpha_2$ dependence of the curves in Fig.~\ref{fig:shat-best} can be attributed to this subtraction. }
We compute $s_{\text{reg}} = \left( P_{\rm reg}\right)_T$ and $n_{\text{reg}} = \left( P_{\rm reg}\right)_\mu$, leading to $(\hat{s}_{\text{reg}})_T \approx -0.30$ and $(\hat{s}_{\text{reg}})_\mu \approx -0.066$ for $(\mu_\crit, T_\crit)$  on Table \ref{table:paras}. Thus,
\begin{align}\label{eq:sr-estimate}
(\hat s_{{\rm reg}})_r  = (\hat s_{{\rm reg}})_T \Delta T_1 - (\hat s_{{\rm reg}})_\mu \Delta \mu_1 \approx 13 \,.,
\end{align}
where we have used the chain rule with $T_r = \Delta T_1$ and $\mu_r = -\Delta \mu_1$, and the values of $\Delta \mu_1$ and $\Delta T_1$ in Table \ref{table:estimation}.

Putting together Eqs.~\eqref{eq:sreg} and~\eqref{eq:sr-estimate} and using $r_{\rm max}$ from Table~\ref{table:estimation}, we find an estimate for the amplitude of the regular part of $\hat s$ at $r=r_{\rm max}$: $\hat{s}_{\rm reg,\,max} \approx -0.01,\, - 0.3$ for Set 1 and 2, respectively. 
The values $|\hat{s}_{\rm reg,\,max}|$ are small compared to $\hat{s}_{\rm max} - \hat{s}_\crit$, shown in Table~\ref{table:estimation}, i.e., our approximation in Eq.~\eqref{eq:shat-approx} is reasonable. Furthermore, we find that the {\em sign} of the regular part, $\hat{s}_{\rm reg}<0$, is the same as that of the energy term contribution, $\shatGrc \tilde \Gr(0) (-r)^{1-\alpha}<0$, in Eq.~(\ref{eq:shat-1st}). Therefore, including the regular term can be seen as a quantitative correction that behaves qualitatively similar to the energy term.

\section{Isentropic trajectories near the QCD critical point}
\label{sec:traj}

In this section, we illustrate the topography of the specific entropy in the $(h,\,r)$ and in the $(\mu,\,T)$ planes around the critical point (not only on the coexistence line).  As discussed in the Introduction, the ideal hydrodynamic evolution of an expanding fireball follows the constant $\hat s$ contour in the $(\mu, T)$ plane. We shall therefore study the family of such isentropic trajectories parameterized by the value of $\hat s$. 

Given the EOS parameters--- the parameters characterizing the location of the critical point $(\mu_c,\, T_\crit,\, s_\crit,\, n_\crit)$ and the mapping parameters $(w,\, \rho,\, \alpha_1,\, \alpha_2)$ --- we use Eq.~(\ref{eq:shat-approx}) to compute $\hat{s}$ over a two-dimensional plane, such as $(r,\,h)$ and $(\mu,\,T)$.

In addition, this analysis uses the expression of $\Gh(h, r)$ and $\Gr(h, r)$ determined by the critical EOS. The derivation of these can be found in Appendix \ref{sec:para}. To evaluate $\hat{s}$ on the $(\mu, T)$ plane, we first solve Eq.~(\ref{eq:X-Y}), leading to
\begin{subequations}
\label{eq:Y(X)}
\begin{align}
h(\mu,\,T) &= - \frac{\Delta\mu \sin \alpha_1 + \Delta T \cos \alpha_1}{T_{\rm c} w \sin \alpha_{12}} \,,\\
r(\mu,\,T) &= \frac{ \Delta\mu \sin \alpha_2 + \Delta T \cos \alpha_2}{T_{\rm c} w \rho  \sin \alpha_{12}}\,.
\end{align}
\end{subequations}
Substituting these into $\Gh(h, r)$ and $\Gr(h, r)$, we can calculate $\Gh(h(\mu, T), r(\mu, T))$ and $\Gr(h(\mu, T), r(\mu, T))$.

In the illustrations below, we use the same Sets 1 and 2 defined in Table \ref{table:paras}. As we observed in the previous section, the values of $\hat s_\crit$ for both Set 1 and 2 given by Eq.~(\ref{eq:shat-set}) and $\cot\alpha_1=14.9$ from Ref.~\cite{Parotto:2018pwx} correspond to the scenario with the specific entropy being nonmonotonic on the low-$T$ (HRG) side of the coexistence line, according to the criterion in Eq.~\eqref{eq:lambda-hill-T}. To illustrate and compare with the scenario where $\hat s$ is nonmonotonic on the QGP side, we consider another alternative value, $\hat s_\crit=12. < \cot\alpha_1$, for both Sets 1 and 2. Thus, for each Set 1 and 2 we now have two alternatives distinguished by the value of $\hat s_\crit$: 
\begin{align}
\label{eq:sc-choices}
\hat s_\crit = 20.3 \,, 19.3\, 
\quad \mbox{or} \quad 12.,
\end{align}
which correspond to ${\rm sgn}\, (\hat s_\crit - \cot \alpha_1) = +1 $ (HRG) and $-1 $ (QGP), respectively.

\begin{figure*}[t]
\centering
\includegraphics[bb=0 0 1520 1450, width=14 cm]{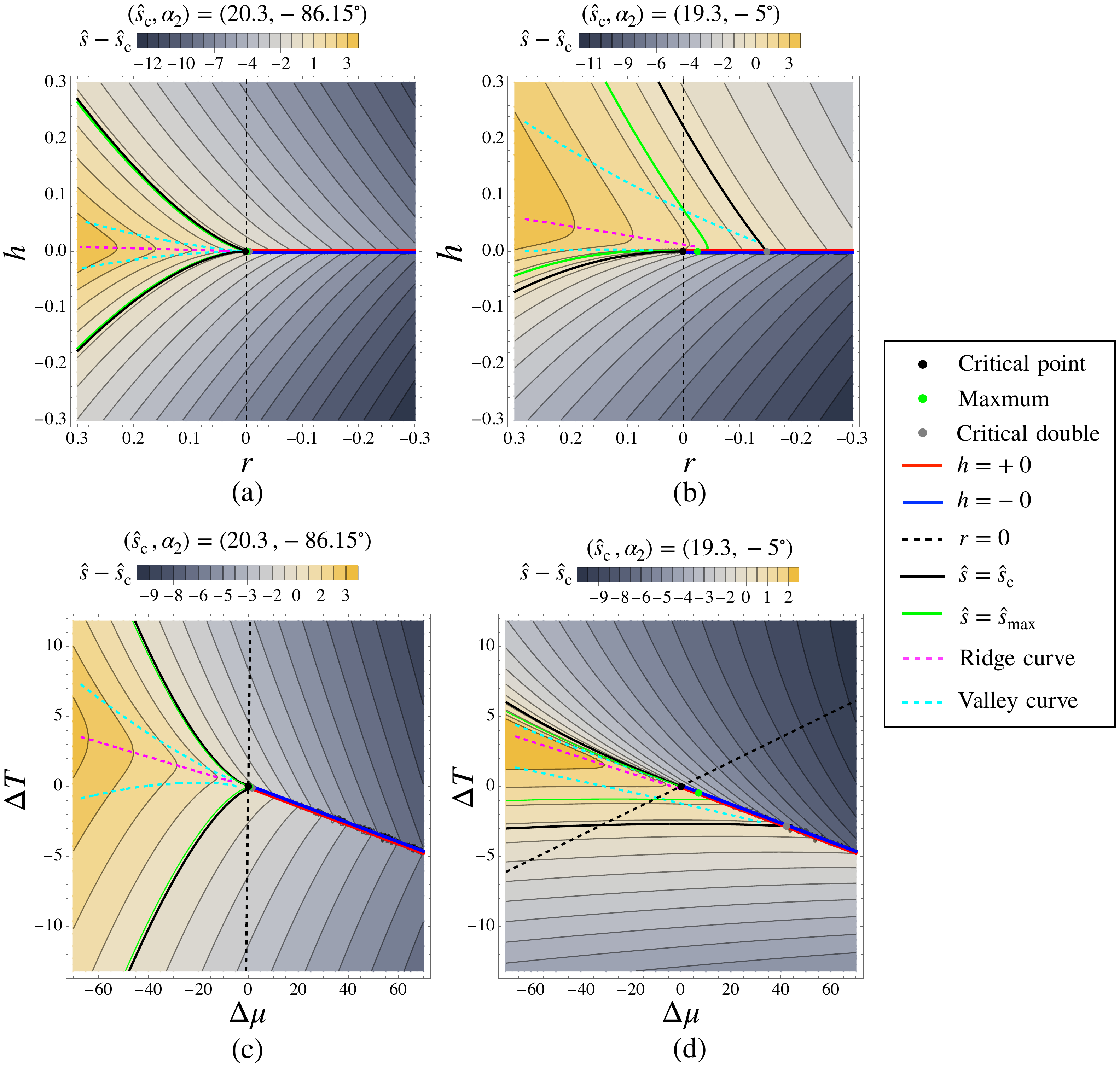}
\caption{
Contours of $\hat s$ computed using Eq.~(\ref{eq:shat-approx}): [(a) and (b)] On $(r,\,h)$ plane; [(c) and (d)] on the $(\mu,\,T)$ plane. We set $\hat s_\crit$ in the first two of Eq.~(\ref{eq:sc-choices}) and use two sets of parameters (see Table \ref{table:paras}) different by the value of $\alpha_2$ shown at the top of each panel. Set 1 is used in panels (a) and (c), while Set 2 -- in (b) and (d). The right legend summarizes the notations of points and curves used in the plots: the black, green, and gray points depict the critical point, the maximum, and the critical double along the nonmonotonic branch as shown in Fig.~\ref{fig:shat-1st}; the red and blue lines and the black dashed lines denote $\hat r$ axis on $h=+0$ and $h=-0$, and the $\hat h$ axis ($r=0$); the black and green solid curves denote the contours with $\hat s=\hat s_\crit$ and $\hat s_{\rm max}$ given by Eq.~(\ref{eq:shat-top}); the magenta and cyan dashed curves show the ridgeline and the valley lines generated based on the mathematical definitions in  Eq.~(\ref{eq:ridge-line-def}). 
The ridge line intersects the coexistence line away from the critical point, even though, due to the smallness of $r_{\rm max}$, it is hard to see, especially for large $\alpha_2$ case [(a) and (c)], as discussed in the main text. The nonmonotonic side appears on the HRG side, according to Eq.~(\ref{eq:lambda-hill-T}), for the present parameter choice.
}
\label{fig:hat_s_contours}
\end{figure*}

\begin{figure*}
\centering
\includegraphics[bb=0 0 1520 1450, width=14 cm]{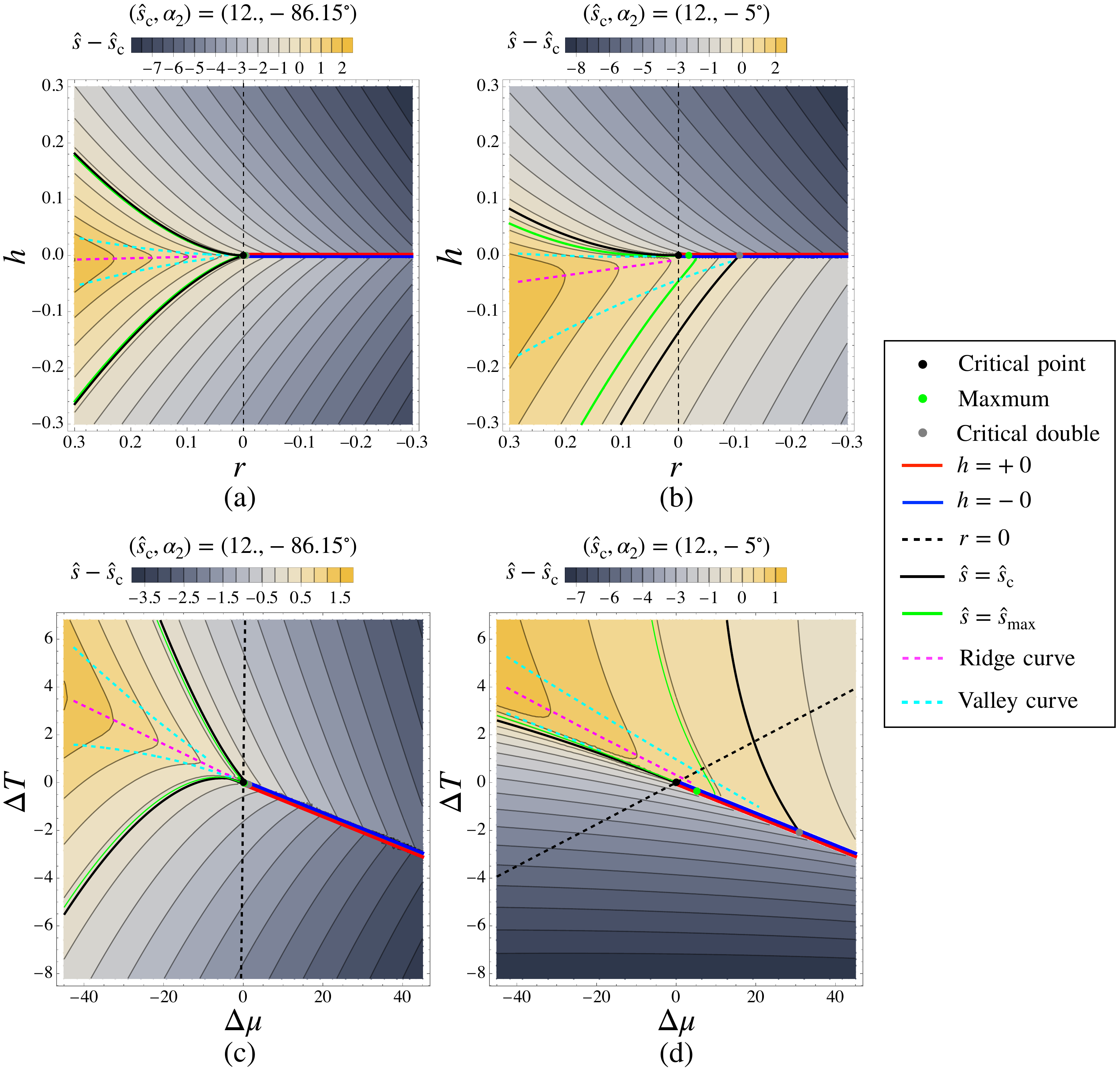}
\caption{Same as Fig.~\ref{fig:hat_s_contours} with the only difference being the choice of $\hat s_c= 12$ from Eq.~(\ref{eq:sc-choices}). Since ${\rm sgn} (\hat s_\crit - \cot \alpha_1) =-1$, the ridge and the nonmonotonic behavior of $\hat s$ along the coexistence line is now on the QGP side, according to Eq.~\eqref{eq:lambda-hill-T}.
}
\label{fig:hat_s_contours-2}
\end{figure*}

We present the contour plots for $\hat s_\crit =20.3\,, 19.3 $ and $\hat s_\crit = 12.$ in Figs.~\ref{fig:hat_s_contours} and \ref{fig:hat_s_contours-2}, respectively. For each set of the parameters shown at the top, we have depicted the contours on two distinct planes, $(r,\,h)$ and $(\mu,\,T)$, represented as panels (a) and (b) as well as (c) and (d), respectively.
All the contours exhibit a distinctive ridge structure. 
We have drawn characteristic ridge lines and valley lines in magenta and cyan dashed lines, respectively, whose mathematical definitions are given by (\ref{eq:ridge-line-def}) (see Appendix~\ref{sec:topo} for details).

The appearance of the nonmonotonic structure along the coexistence line can be seen as a consequence of the coexistence line cutting across the ridge of $\hat s$.
The mixing between $\Gh$ and $\Gr$ in Eq.~(\ref{eq:shat-approx}) for $\hat s$ modifies the ridgeline of $\Gr(h,\,r)$ from its symmetric location along the crossover line $h=0$ shown in  Fig.~\ref{fig:m-sigma-R-theta}\,(d). The ridge line for $\hat s$ is bent and shifted along the coexistence line, intersecting it away from the critical point, at $r=r_{\rm max}$.
Since $|r_{\rm max}|$ is small (see Table \ref{table:estimation}), the nonmonotonic structure (the maximum and the critical double) is squeezed toward the critical point, especially for Set 1 with large $\alpha_2$ (see Sec.~\ref{sec:lattice} for explanation).
It is evident that the nonmonotonic shape observed along the coexistence line in Fig.~\ref{fig:shat-1st} is a cross section of the ridge.

\section{Slope formula and the classification of the contours}
\label{sec:slope}

\subsection{The slope formula}
\label{sec:slope-formula}

To closely examine the contours near the coexistence line, we calculate the slope of the fixed $\hat s$ contours, expressed as
\begin{align}
\label{eq:sl_def}
\tan\alpha_{\hat s}\equiv -\cx{\left( \frac{\partial T}{\partial \mu} \right)_{\hat s} } =  \cx{\frac{ \hat s_\mu }{\hat s_T}} \,.
\end{align}
We compute the ratio in Eq.~\eqref{eq:sl_def} using
\begin{subequations}
\begin{align}
\label{eq:s1st-pm-dr}
\cx{\hat{s}_r} & = 
(-r)^{\beta-1}\left[
\shatGhc \Ghrtil(\pm0) + \shatGrc \Grrtil(0) x 
\right]\,, \\
\label{eq:s1st-pm-dh}
\cx{\hat{s}_h} & = (-r)^{-\gamma}\left[ \shatGhc \Ghhtil(0) + \shatGrc \Grhtil(\pm0) x \right] \,,
\end{align}    
\end{subequations}
and Eqs.~\eqref{eq:Y_X}. Here, we used $\beta - 1 + \gamma = 1-\alpha-\beta$ and introduced 
\begin{equation}
    x \equiv (-r)^{1-\alpha-\beta},
\end{equation}
and the critical amplitudes, such as $\Ghrtil(\pm0)$, defined in Appendix~\ref{sec:susceptibilities}.

We consider the angle of the slope $\alpha_{\hat s}$ relative to that of the coexistence boundary, $\alpha_1$, and obtain the following formula: 
\begin{align}
\label{eq:slope-formula-alpha}
&\tan \left( \cx{\alpha_{\hat s}} -\alpha_1\right) 
={x(x \mp  x_{\rm max})\sin\alpha_{12}} \notag\\
&\times\Bigg[\left(
x_{\rm max}\frac{\Ghhtil(0)}{\beta} \mp x\frac{\beta}{\Grrtil(0)}\right)\rho
- x(x\mp x_{\rm max})\cos\alpha_{12}\Bigg]^{-1}\,, 
\end{align}
where $\Ghhtil(0)>0$ and $\Grrtil(0)>0$ and we introduced a signed quantity,
\begin{equation}\label{eq:xmax}
    x_{\rm max}\equiv {\rm sgn}(\hat s_\crit - \cot\alpha_1)(-r_{\rm max})^{1-\alpha-\beta}\,, 
\end{equation}
which carries information about how far from the critical point the maximum occurs [see Eqs.~\eqref{eq:rmax} and \eqref{eq:rmax-alpha}] and also about which side of the coexistence curve is nonmonotonic [see Eq.~\eqref{eq:lambda-hill-T}].
As before, $\pm$ in 
Eq.~\eqref{eq:slope-formula-alpha} correspond to $h\to\pm0$ sides of the coexistence line, respectively.

\begin{figure}[t]
\centering
\includegraphics[bb=0 0 580 420, width=8 cm]{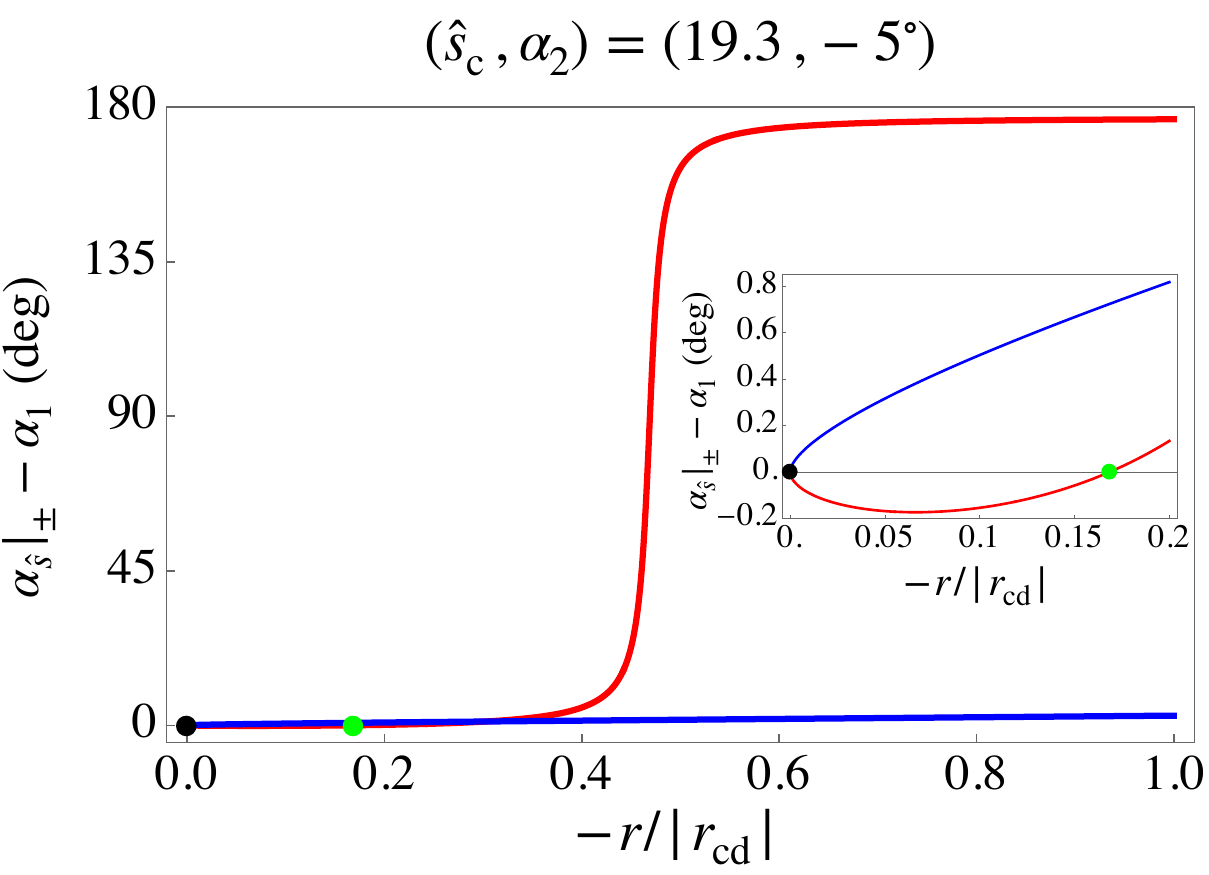}
\caption{The angle between an isentropic curve and the coexistence line at their intersection as a function of the distance of the intersection point from the critical point. 
The angle is found from Eq.~(\ref{eq:slope-formula-alpha}) and plotted with $h\to+0$ branch in red and $h\to -0$ in blue for the choice of parameters corresponding to Set 2 in Table \ref{table:paras} (small $\alpha_{12}$) and $x_{\rm max}>0$ in Eq.~\eqref{eq:xmax}. The inset shows the nonmonotonic structure near the critical point.
}
\label{fig:slope-angle}
\end{figure}

Figure~\ref{fig:slope-angle} illustrates the relative angle computed by (\ref{eq:slope-formula-alpha}) on each side of the coexistence line for Set 2.
The angle $\alpha_{\hat s}-\alpha_1$ is nonmonotonic on the branch where $\hat s$ is nonmonotonic. 
At the maximum of $\hat s$, at $x=x_{\rm max}$, the relative angle vanishes. In other words, $\hat s =\hat s_{\rm max}$ contour is tangential to the coexistence line. For $x>x_{\rm max}$ the relative angle increases, crossing 90$^\circ$ at the point where the denominator in Eq.~\eqref{eq:slope-formula-alpha} vanishes. At this point, the isentrope is perpendicular to the coexistence line.

\subsection{A cusp for small $|r_{\rm max}|$}

From the examples in Figs.~\ref{fig:hat_s_contours}(a) and \ref{fig:hat_s_contours}(c) and Figs.~\ref{fig:hat_s_contours-2}(a) and \ref{fig:hat_s_contours-2}(c) (Set 1 parameters), we see that $|r_{\rm max}|$ is so small that the resolution of the phase diagram is not sufficient to distinguish points $r=0$, $r_{\rm max}$, and $r_{\rm cd}$, characterizing the nonmonotonic structure. There appear to be two isentropes (solid black) meeting at the critical point and forming a cusp. 
Such a cusp has indeed been observed in Refs.~\cite{ Parotto:2018pwx,Dore:2022qyz}
and subsequent literature, where  Set 1 was used, as well as earlier, in Ref.~\cite{Nonaka:2004pg}, where $\alpha_{12}= 90^\circ$ choice was made.

We can understand the origin of this cusp if we remember that small $|r_{\rm max}|$ is a consequence of $\shatGhc\ll\shatGrc$. In this case, neglecting the first term in the expansion of $\hat s$ in Eq.~\eqref{eq:shat-approx}, we conclude that the critical isentrope corresponds to constant $\Gr$:
\begin{align}
\label{eq:crit-isentrope-default}
\hat s = \hat s_\crit \quad \Leftrightarrow \quad \Gr = 0\,.
\end{align}
The corresponding contour is defined in the $(h,\,r)$ space by $z=z_\Gr$, where $\tilde\Gr(z_\Gr)=0$.%
\footnote{
In terms of $\theta_\varepsilon$ discussed below Eq. (\ref{eq:ceps}),
$z_\varepsilon = \bar h(\theta_\varepsilon)/(1-\theta_\varepsilon^2)^{\beta\delta}$.
}
Such a constant $z$ contour has a cusp since it is given by $h=z_\Gr |r|^{\beta\delta}$ with $\beta\delta>1$.

In the next subsection, we shall zoom into the region $0>r>r_{\rm cd}$ to study how the nonmonotonic structure there affects the isentropes in more detail.

\subsection{Classification of trajectories}
\label{sec:class}

\begin{figure}[t]
\centering
\includegraphics[bb=0 0 710 430, width=8.5 cm]{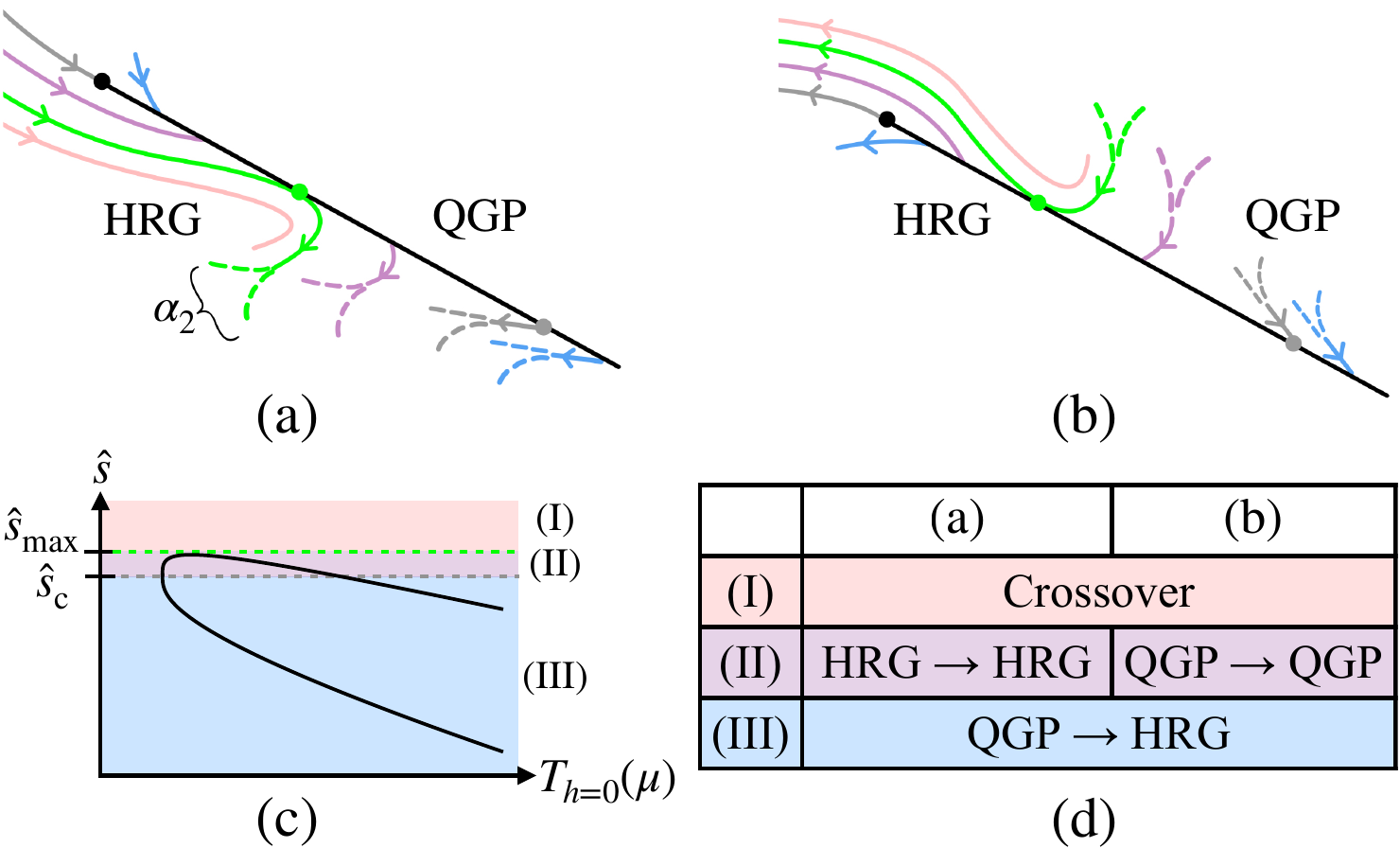}
\caption{Classification of isentropic trajectories near the QCD critical point. 
In panel (a) the maximum of $\hat s$ is on the HRG side, while in panel (b) it is on the QGP side. The arrows point in the direction the entropy density decreases (the system expands). 
The gray curve is the critical isentrope ($\hat s = \hat s_\crit$). The contour $\hat s = \hat s_{\rm max}$ given by Eq.~(\ref{eq:shat-top}) is the green curve. The pink, purple, and blue curves are representative contours in each range of $\hat s$ shown in the corresponding color in panels (c) and (d). The dashed lines in (a) and (b) express details depending on the choice of $\alpha_2$, which can be seen as a difference between, e.g.,  Figs.~\ref{fig:hat_s_contours}(c) and \ref{fig:hat_s_contours}(d), away from the coexistence line.
Panel (d) summarizes the classification based on the value of $\hat s$ (see text).}
\label{fig:class_contours}
\end{figure}

We represent different types of isentropic trajectories near the coexistence line schematically in Fig.~\ref{fig:class_contours}. Figures \ref{fig:class_contours}(a) and \ref{fig:class_contours}(b) correspond to the scenarios where the nonmonotonicity appears on the HRG and QGP side, ${\rm sgn} (\hat s_\crit - \cot \alpha_1 )=+1$ and $-1$, respectively. These sample trajectories intersect the coexistence line at a relative angle $\alpha_{\hat s}-\alpha_1$ in agreement with Eq.~(\ref{eq:slope-formula-alpha}) (see also Fig.~\ref{fig:slope-angle} for the nonmonotonic HRG side scenario)

We can classify the isentropic trajectories based on their initial specific entropy, $\hat s=\hat s_{\rm ini}$. Each trajectory falls into one of the following three families depending on the relationship between $\hat s_{\rm ini}$ and the two characteristic values:
$\hat s_\crit$ and $\hat s_{\rm max}$ given by Eq.~(\ref{eq:shat-top}):

(I) Trajectories with $\hat s_{\rm ini}>\hat s_{\rm max}$  do not encounter the first-order transition and pass through the crossover region. These are shown as pink curves in Figs.~\ref{fig:class_contours}(a) and \ref{fig:class_contours}(b), corresponding to horizontal lines in the pink area in Fig.~\ref{fig:class_contours}(c);

(II) Trajectories with $
\hat s_\crit < \hat s_{\rm ini} < \hat s_{\rm max}$ enter the coexistence region from one side (the nonmonotonic one) and exit on the {\em same\/} side. These are shown as purple curves in Figs.~\ref{fig:class_contours}(a) and \ref{fig:class_contours}(b), and correspond to horizontal lines in the purple region in Fig.~\ref{fig:class_contours}(c);

(III) Trajectories with $
\hat s_{\rm ini} < \hat s_\crit $ enter the coexistence region from one side and exit on the opposite side [the blue curve in Figs.~\ref{fig:class_contours}(a) and \ref{fig:class_contours}(b), and the blue region in Fig.~\ref{fig:class_contours}(c)].

The same three classes of trajectories exist in both scenarios distinguished by the location of the specific entropy maximum  -- HRG or QGP side, 
as illustrated in Figs.~\ref{fig:class_contours}(a) and \ref{fig:class_contours}(b), respectively. 
The main difference between the two scenarios is the direction of the $T$ (and $\mu$) change on crossing the coexistence line. Under the ``HRG-side scenario'' the temperature decreases, 
while under the ``QGP-side scenario''  the temperature increases.  
Both scenarios have been discussed in the literature (see, e.g., Refs.~\cite{Stephanov:1998dy,Scavenius:2000qd,Subramanian:1986xh}).

The most notable are the class II trajectories, entering and exiting the coexistence region on the same side. The detailed dynamics of the transition are beyond the scope of this paper, which only discusses adiabatic trajectories. However, it is worth noting that even if non-negligible entropy is produced during the traversal of the coexistence region, the additional entropy will only move the exit point of the class II trajectory closer to the maximum of $s/n$, but still on the same side of the transition.

\section{Conclusion}
\label{sec:dis}

In this paper, we set out to understand the behavior of the isentropic trajectories on the QCD phase diagram in the $(\mu,\,T)$ plane in the presence of the QCD critical point. The hydrodynamic expansion of a heavy-ion collision fireball follows such trajectories as far as the entropy production is negligible, i.e., specific entropy $s/n$ is conserved, which is a reasonable and often used approximation.

The main observation of this paper is that the specific entropy as a function of the distance from the critical point along the coexistence line must be nonmonotonic, and specifically must exhibit a maximum. This is a robust and model-independent consequence of the following two facts. First, the universal critical behavior dictates that the specific entropy must {\em rise} on one of the two sides of the coexistence line as a function of $|T-T_c|$. Second, the third law of thermodynamics dictates that the specific entropy must {\em fall} to zero at $T=0$.

We find that the maximum could occur on either side of the coexistence line: high-$T$ (QGP) or low-$T$ (HRG) side. We find a criterion that determines which side is nonmonotonic. It is determined by the sign of  $\hat s_\crit-\cot\alpha_1$, i.e., the difference between the specific entropy and the inverse slope of the coexistence line, both quantities evaluated at the critical point.

When this discriminant quantity is sufficiently small, the maximum occurs close to the critical point, where the equation of state is determined universally, up to a small number of nonuniversal parameters standardized by Ref.~\cite{Parotto:2018pwx}. We use this regime to demonstrate analytically how the maximum moves as a function of the critical point parameters.

We also show that for the critical EOS parameters discussed in Ref.~\cite{Parotto:2018pwx}, and constrained by lattice data, the discriminant parameter $\hat s_\crit-\cot\alpha_1$ is positive, which according to Eq.~\eqref{eq:lambda-hill-T} means that the maximum is on the HRG side of the coexistence line. Furthermore, we show the maximum occurs in the regime where the EOS is universal, i.e., dominated by the critical singularity.

Turning to the behavior of the isentropic trajectories in the ($\mu$,\,$T$) plane, we demonstrate that the maximum on the coexistence line can be viewed as a cross section of a ``ridge'' of $s/n$ ``landscape,'' where 
the trajectories are contours or equal elevation lines. The ridge topography also helps explain the critical ``focusing" or "lensing" effects observed in Refs.~\cite{Nonaka:2004pg,Dore:2022qyz}.

We classify the trajectories, or contours, according to how they cross (if they do) the coexistence line, which represents a ``cliff'' in the landscape.
The most unusual trajectories are such that the fireball enters and then exits the coexistence region on the {\em same} side.

Incidentally, the same discriminant parameter $\hat s_c-\cot\alpha_1$, or $(dP/dT)_n$ [see Eq.~\eqref{eq:dPdT}], determines whether the temperature is higher or lower after the transition. For the scenario where the maximum is on the HRG side, the temperature is lower.

It must be emphasized, that we do not attempt to describe the evolution inside the coexistence region, but only follow the commonly used assumption that the entropy production is negligible. It would be interesting to consider the effect of entropy production, but this should be done within a more detailed dynamical description of the first-order transition. While such a study is beyond the scope of this paper, we hope that our findings will help to advance the understanding of the first-order transition dynamics in heavy-ion collisions.

\begin{acknowledgements}
N.S. thanks, Xin An and Masaru Hongo, for useful comments and discussion. This work is supported by the U.S. Department of Energy, Office of Science, Office of Nuclear Physics Award No. DE-FG0201ER41195. M.P. is supported by the U.S. Department of Energy, Office of Nuclear Physics under Award No. DE-FG02-93ER40762.
\end{acknowledgements}

\appendix

\section{Susceptibilities in a scaling form}
\label{sec:susceptibilities}
In this Appendix, we summarize the susceptibilities used in Secs.~\ref{sec:s/n-first} and \ref{sec:slope}.

Using Eq.~\eqref{eq:m-sigma-sca}, we can write all four susceptibilities in scaling forms of the $z=h/(-r)^{\beta\delta}$ variable (for $r<0$):
\begin{subequations}
\label{eq:suscept-h0}
\begin{align}
\label{eq:suscept}
{\Gh_h} & = G_{hh} = {\Ghhtil}(z) (-r)^{-\gamma} \,,\\
\label{eq:suscept-cross-1}
{\Gh_r} & = G_{hr} = {\Ghrtil(z)} (-r)^{\beta-1} \,, \\
\label{eq:suscept-cross-2}
{\Gr_h} & = G_{rh} = {\Grhtil(z)} (-r)^{\beta-1}  \,,\\
\label{eq:suscept-2}
{\Gr_r} & = G_{rr} = {\Grrtil(z)} (-r)^{-\alpha} \,,
\end{align}
\end{subequations}
with $\gamma = \beta(\delta - 1)$ and%
\footnote{We shall numerically calculate, using the well-known parametric representation of the EOS, $\tilde \Gh(z)$, $\tilde \Gh'(z)$, $\tilde \Gr(z)$, and $\tilde \Gr'(z)$ for $r\leq0$ in Eqs.~(\ref{eq:m-sca-def-z-2}), (\ref{eq:sca-m-der}), (\ref{eq:sigma-sca-def-z-2}), and (\ref{eq:eps-til-z-der}), respectively. The referred equations, which are also valid for $r>0$ branches, are labeled by the sign of $r$ (see also Fig.~\ref{fig:scaling-z} for illustration).}
\begin{subequations}
\label{eq:sus-sca-til}
\begin{align}
\Ghhtil(z) &= \tilde \Gh'(z)\,, \\
\Ghrtil(z) &= -\beta \tilde \Gh(z)\,,\\
\Grhtil(z) &= \tilde \Gr'(z)\,, \\
\Grrtil(z) &= -(1-\alpha) \tilde \Gr(z) \,.
\end{align}
\end{subequations}
Here, the prime symbol indicates differentiation with respect to the function's argument, i.e., $z$.
Among the four susceptibilities, $\Gh_h$ and $\Gr_r$ represent the magnetic susceptibility and the specific heat (at constant $h$), respectively, while $\Gh_r$ and $\Gr_h$ denote cross susceptibilities. Note that thermodynamics stability requires $\Gh_h\geq0$, $\Gr_r\geq0$, and $(\Gh_h \Gr_r - \Gh_r \Gr_h)\geq0$. 

For our discussion along the coexistence line in Secs.~\ref{sec:s/n-first} and \ref{sec:slope}, we need the values of the scaling functions (\ref{eq:sus-sca-til}) at $h=\pm0$, or, equivalently, $z=\pm 0$:%
\footnote{
These coefficients are known as critical amplitudes and in the common notation (see, e.g., Ref.~\cite{Pelissetto:2000ek}): $\Grrtil(0)=A^-$, $|\Grhtil(0)|=\beta B$, and  $\Ghhtil(0)=C^-$ (where ``$-$'' refers to $r<0$).
}
\begin{subequations}
\label{eq:suscept-cofs}
\begin{align}
{\Ghhtil(0)} &= \tilde \Gh'(0) \,,\\
{\Ghrtil(\pm0)} &= \mp\beta \,, \\
{\Grhtil(\pm0)} &= \tilde \Gr'(\pm 0)\,,\\ 
\label{eq:G-tilde-rr}
{\Grrtil(0)} &= -(1-\alpha) \tilde \Gr(0) \,.
\end{align}
\end{subequations}
The cross susceptibility incorporates the sign of $h$, whereas the magnetic susceptibility and the specific heat are even under $h\rightarrow -h$.

Among four values of (\ref{eq:suscept-cofs}), only two are unknown. This is because the normalization of $\tilde \Gh(\pm0)=\pm1$ have fixed one of them and the Maxwell relation,
\begin{align}
\label{eq:maxwell}
\Gh_r = \Gr_h\,,
\end{align}
yields $\tilde \Gr'(\pm 0) = \mp \beta $.

For a detailed evaluation of the unknown values, we refer to Appendix \ref{sec:s-1st-der}, specifically, Eqs.~(\ref{eq:m-sca-slope-z0}) and (\ref{eq:sigma_m_z=0}):
\begin{align}
\label{eq:para-comp}
\tilde\Gh'(0) \equiv
\tilde\Gh'_-(0)\approx 0.35 \,, \quad \tilde \Gr(0) \equiv
\tilde\Gr_-(0)\approx -0.66\,.
\end{align}
where subscript ``$-$'' omitted in Eq.~\eqref{eq:m-sigma-sca} refers to $r<0$.

\section{3D Ising equation of state}
\label{sec:para}
This Appendix is a review of the 3D Ising EOS. Specifically, we show how to obtain $\Gh(h,\,r)$ and $\Gr(h,\,r)$ numerically, used in Sec.~\ref{sec:traj}. A summary can be found at the end.

We employ a parametric representation of the EOS that is valid near an arbitrary critical point belonging to the 3D Ising universality class \cite{Schofield:1969zza,Guida:1996ep}, e.g., the liquid-gas and QCD critical points, etc. The relevant parameters of the renormalization group, namely the external field $h$ and the reduced temperature $r$, denoted as $Y \equiv (h,\,r)$, can be transformed into coordinates $Z \equiv (\theta,R)$ using the following relations:
\begin{subequations}
\label{eq:r,h}
\begin{align}
\label{eq:h}
h &= R^{\beta \delta}\bar  h (\theta)\,,\\
\label{eq:r}
r &= R(1-\theta^2) \,,
\end{align}
\end{subequations}
where $R\geq0$ and the magnetization $\Gh$ is expressed as:
\begin{align}
\label{eq:m}
\Gh & = R^\beta \bar{\Gh}(\theta) \,.
\end{align}
We use the scaling forms of $\Gh$ and $h$ as functions of $\theta$, denoted as $\bar {\Gh}(\theta)$ and $\bar  h(\theta)$, respectively:
\begin{subequations}
\label{eq:m_h_tils}
\begin{align}
\label{eq:m_h_tils-1}
\bar \Gh (\theta) &= \Gh_0 \theta \,, \\
\label{eq:m_h_tils-2}
\bar h (\theta) &= h_0 \theta (1 + a \theta^2 + b \theta^4)\,,
\end{align}
\end{subequations}
where the coefficients $a \approx -0.762$ and $b \approx 0.008$ are obtained through renormalization group analysis using a nonperturbative resummation \cite{Guida:1996ep,Zinn-Justin:1998qip}. To distinguish the scaling functions of $z$ introduced in Eq.~(\ref{eq:m-sigma-sca}), we use the bar notation for the scaling form of $\theta$. 
The normalization constants $\Gh_0$ and $h_0$ are fixed by the conditions $1=\Gh(h=+0,r=-1)=\Gh(h=1,r=0)$, which reduces to
\begin{subequations}
\label{eq:norm}
\begin{align}
\label{eq:norm-1}
1 &= (\theta_{\rm max}^2-1)^{-\beta} \Gh_0 \theta_{\rm max} \,, \\
1 &= \Gh_0^{-\delta} h_0 (1+a+b)\,,
\end{align}
\end{subequations}
where $\theta_{\rm max} \simeq 1.15$ is a nontrivial root of $\bar  h (\theta)=0$. By solving Eq.~(\ref{eq:norm}) with the parameter values, $a,b$, etc., we obtain $\Gh_0\approx 0.60$ and $h_0\approx 0.36$.%
\footnote{These values are the same as the literature, e.g., Ref.~\cite{Mroczek:2022oga}.}
One of the benefits of working in $Z$ coordinate is that the relation between $\Gh$ and $h$ is well organized as Eq.~(\ref{eq:m_h_tils}).

\begin{figure*}[t]
\centering
 \includegraphics[bb=0 0 1160 1320, width=14 cm]{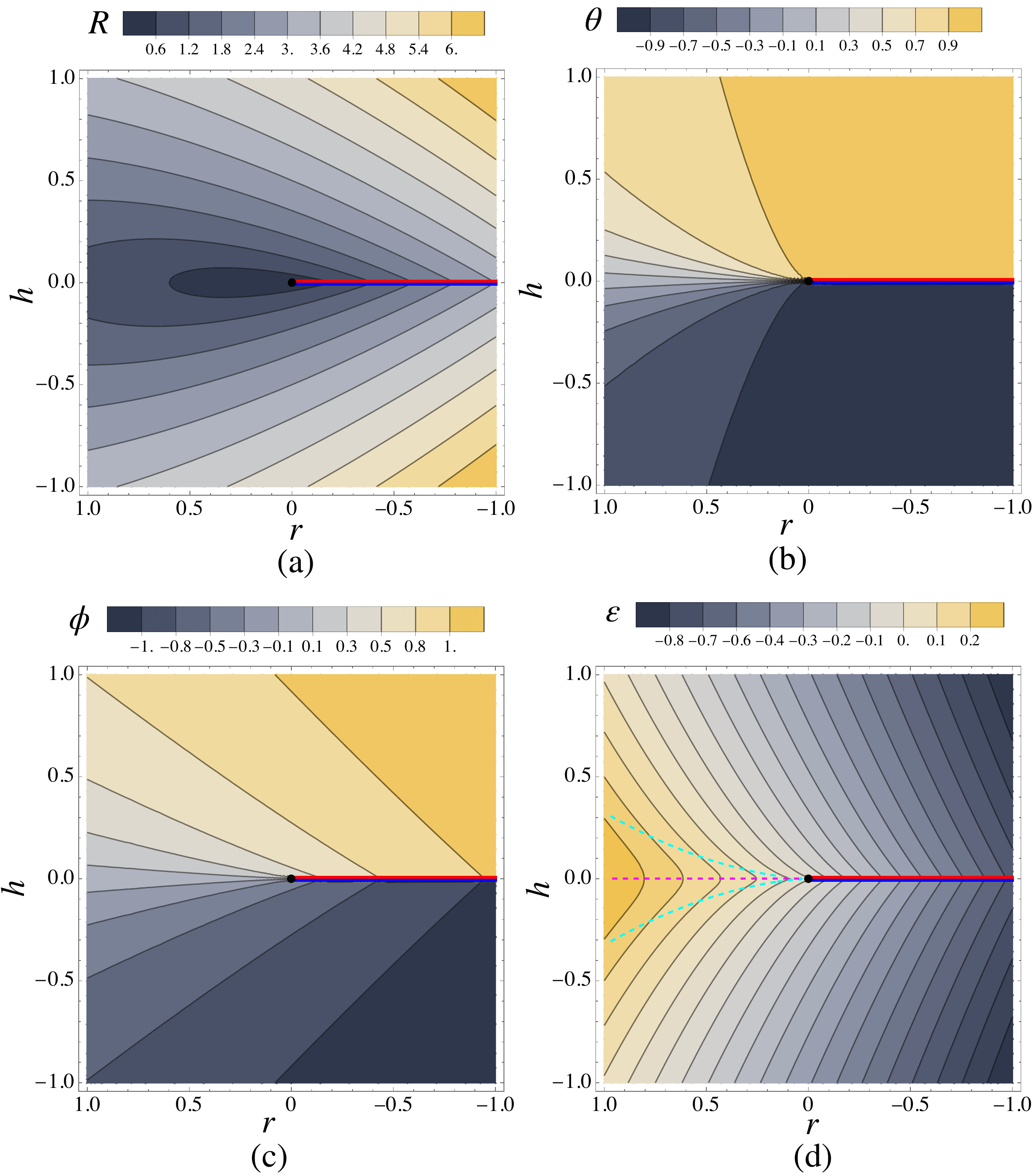}
\caption{Contours on the $(r,\,h)$ plane with fixed (a) $R$, (b) $\theta$, (c) $\Gh$, and (b) $\Gr$. Red ($\theta=\theta_{\rm max}$) and blue ($\theta=-\theta_{\rm max}$) lines at $r<0$ represent each side separated by the first-order boundary. The magenta and cyan dashed lines on (d) are the ridgeline and the valley lines whose mathematical definition will be given by (\ref{eq:ridge-line-def}).}
\label{fig:m-sigma-R-theta}
\end{figure*}

We numerically solve Eq.~(\ref{eq:r,h}) over a specific closed range of $Y$, yielding $\theta(h,\,r)$ and $R(h,\,r)$ denoted by $Z(Y)$. Then, we substitute $Z(Y)$ into (\ref{eq:m}) to obtain $\Gh(Y)$. In Figs.~\ref{fig:m-sigma-R-theta}(a)--\ref{fig:m-sigma-R-theta}(c), we present the contours of $R$, $\theta$, and $\Gh$, respectively. As observed from (a) and (b), $R$ ($R\geq 0$) represents the distance from the critical point at the origin, while $\theta$ ($-\theta_{\rm max} \leq \theta \leq \theta_{\rm max})$ measures the distance along the constant $R$ contour from the edges of the first-order boundary ($h=0,\ r<0$). Here, $\theta = \pm \theta_{\rm max}$ corresponds to $h=\pm 0$. With this parametrization, we can describe the criticality as a function of the $R$ variable, and the thermodynamic quantities will exhibit smooth behavior in the $\theta$ variable.

The critical part of the thermodynamic pressure $p$ (or the Gibbs free energy, $p=G$) also follows a scaling form:
\begin{align}
\label{eq:p}
p = R^{2-\alpha} \bar  p (\theta)\,.
\end{align}
We will determine the scaling function $\bar p(\theta)$ below to match with the relation between $\Gh$ and $h$, Eq.~(\ref{eq:m_h_tils}). With Eq.~(\ref{eq:p}) as a starting point, we can calculate the equations for the magnetization $\Gh$ and the Ising energy $\Gr$ as (\ref{eq:GhGr}):
\begin{subequations}
\label{eq:m-sigma}
\begin{align}
\label{eq:m-p_expr}
\Gh & = R^{\beta} \left[ 2(2-\alpha)\theta \bar {p}(\theta)+ \left(1-\theta^2 \right) \bar {p}'(\theta) \right] /\bar  J(\theta)  \,,\\
\label{eq:sigma_til-def}
\Gr & = R^{1-\alpha} \left[(2-\alpha)\bar  h'(\theta) \bar {p}(\theta)-\beta \delta \bar  h(\theta) \bar {p}'(\theta)\right]/\bar  J(\theta)\,.
\end{align}
\end{subequations}
To derive the right-hand sides of Eq.~(\ref{eq:m-sigma}), we use the derivative formula for an arbitrary function, $f\equiv R^{a_f} \bar  f(\theta)$, with the scaling dimension $a_f$:
\begin{align}
\label{eq:ｄel_A}
f_Y & \equiv \partial_Y f = R^{a_f-1}\left[R \bar {f}'(\theta)\theta_{Y}+a_f \bar {f}(\theta)R_{Y}\right]\,,
\end{align}
where $Z_Y$ is the following inverse of Jacobian matrix $\mathcal J = Y_Z$:
\begin{align}
\label{eq:inv-jaco}
\left(
\begin{array}{cc}
\theta_h & \theta_r \\
R_h & R_r 
\end{array} 
\right) = \frac{1}{\bar  J(\theta)}
\left(
\begin{array}{cc}
R^{-\beta \delta} \left(1-\theta^2 \right) & -R^{-1} \beta \delta \bar  h(\theta) \\
2 R^{1-\beta  \delta } \theta & \bar {h}'
(\theta)
\end{array}
\right)\,,
\end{align}
with the Jacobian $J\equiv|\mathcal{J}|= R^{\beta \delta} \bar  J(\theta)$ and its scaling form:
\begin{align}
\bar  J(\theta) &= (1-\theta^2) \bar {h}' (\theta) + 2 \beta \delta \theta \bar  h (\theta) \,.
\end{align}

We combine the original definition of $\Gh$ (\ref{eq:m}) and its relation to $\bar  p(\theta)$ (\ref{eq:m-p_expr}), yielding
\begin{align}
2(2-\alpha)\theta \bar {p}(\theta) + (1-\theta^2) \bar {p}'(\theta) = \bar  {\Gh} (\theta) \bar  J(\theta)\,,
\end{align}
which determines the form of $\bar  p(\theta)$. Given that this is a first-order nonhomogeneous equation, it includes both special and general solutions. The special solution is relevant to our current analysis concerning the critical exponents beyond the mean-field level: $\bar p (\theta) = \sum_{n=0}^3 C_n (1-\theta^2)^n$, where $C_n$ are functions of the critical exponents and the parameters in Eq.~(\ref{eq:m_h_tils}) (for those explicit forms, refer to Ref.~\cite{Parotto:2018pwx}). The constant of integration associated with the general solution $\propto (\theta^2-1)^{2-\alpha}$ is determined to be zero so that any singularity does not arise in the scaling part of the pressure, $\bar p(\theta)$.

We substitute the solution $\bar  p  (\theta)$ into its relation to the Ising energy density (\ref{eq:sigma_til-def}), and obtain $\Gr$ with its explicit scaling form: 
\begin{subequations}
\label{eq:sigma-expr}
\begin{align}
\label{eq:sigma-R-theta}
\Gr &=R^{1-\alpha} \bar \Gr (\theta)\,,\\
\label{eq:sigma-theta}
\bar \Gr (\theta) &= \Gh_0 h_0 \left[ c_{\Gr 0} + c_{\Gr 1} (1-\theta^2) + c_{\Gr 2} (1-\theta^2)^2  \right] \,,
\end{align}
\end{subequations}
with 
\begin{subequations}\label{eq:ceps}
\begin{align}
c_{\Gr 0} &= \frac{-(\alpha -2) (a+b+1)-2 \beta  (2 a+3 b+1)}{2 (\alpha -1)} \,,\\
c_{\Gr 1} &= \frac{(\alpha -2) (a+2 b)+4 \beta  (a+3 b)}{2 \alpha }\,,\\
c_{\Gr 2} &= -\frac{b (\alpha +6 \beta -2)}{2 (\alpha +1)} \,.
\end{align}
\end{subequations}
Using the actual values of $\alpha,\beta,a$, and $b$ we can estimate the coefficients  as $(c_{\Gr 0},c_{\Gr 1},c_{\Gr 2}) \approx (-0.4,\, 2.0,\, -2 \times 10^{-4})$. Thus, $\bar \Gr(\theta)$ is roughly a convex parabola whose maximum and zero points are located at $\theta=0$ and $\theta = \pm \theta_\Gr\approx\pm\, 0.88$, respectively.

We highlight the qualitative distinction in behaviors for $\Gh$ and $\Gr$, in Figs.~\ref{fig:m-sigma-R-theta}(c) and \ref{fig:m-sigma-R-theta}(d), respectively. The former exhibits a {\it monotonic} increase as $\theta$ rises while keeping $R$ constant [also refer to Figs.~\ref{fig:m-sigma-R-theta}(a) and \ref{fig:m-sigma-R-theta}(b) for tracking $R$ and $\theta$]. In contrast, the convex nature of $\Gr(\theta)$ results in a {\it nonmonotonic} ridge structure of $\Gr$. The distinctive ridgeline and the valley lines are illustrated in the magenta and cyan dashed lines according to the mathematical definition given by Eq.~(\ref{eq:ridge-line-def}). Specifically, the ridgelines form a straight line with $\theta =0$ parametrized by an arbitrary $R>0$, due to symemtry of $\Gr$ under $h\rightarrow -h$.%
\footnote{This is in contrast to $\hat s$ as we discuss at the end of Sec.~\ref{sec:traj}.}

We shall also compute the susceptibilities in the $(R,\theta)$ scaling form. By applying the derivative formula Eq.~(\ref{eq:ｄel_A}) to $\Gh$ and $\Gr$ given by (\ref{eq:m}) and (\ref{eq:sigma-R-theta}), respectively, we obtain
\begin{subequations}
\label{eq:sus}
\begin{align}
\label{eq:sus-ch1}
\Gh_h 
&= R^{-\gamma}\left[2 \beta \theta \bar {\Gh}(\theta)+ \left(1-\theta^2 \right) \bar {\Gh}'(\theta)\right]/\bar  J(\theta) \,,\\
\label{eq:sus-cr1}
\Gh_r 
&=R^{\beta-1}\left[\beta \bar {h}'(\theta) \bar {\Gh}(\theta)-\beta \delta \bar  h(\theta) \bar {\Gh}'(\theta)\right]/\bar  J(\theta) \,,\\
\label{eq:sus-cr2}
\Gr_h
&= R^{\beta-1}\left[2 (1-\alpha) \theta \bar {\Gr}(\theta)+ \left(1-\theta^2 \right) \bar {\Gr}'(\theta)\right]/\bar  J(\theta) \,,\\
\Gr_r 
&=R^{-\alpha}\left[(1-\alpha) \bar {h}'(\theta) \bar {\Gr}(\theta)-\beta \delta \bar  h(\theta) \bar {\Gr}'(\theta)\right]/\bar  J(\theta)  \,.
\end{align}
\end{subequations}
One can check that the the Maxwell relation (\ref{eq:maxwell}) holds for the expressions (\ref{eq:sus-cr1}) and (\ref{eq:sus-cr2}), using Eq.~(\ref{eq:m-sigma}).

In summary, we determined \(\Gh(h, r)\) and \(\Gr(h, r)\) using the auxiliary variables \((\theta, R)\) through the following steps:
\begin{enumerate}
\item We derived analytic expressions for $\Gh(\theta, R)$, Eqs.~(\ref{eq:m}) and (\ref{eq:m_h_tils-1}), and $\Gr(\theta, R)$, (\ref{eq:sigma-expr}), respectively.
\item We numerically solved Eq.~(\ref{eq:r,h}) with (\ref{eq:m_h_tils-2}), within a specified range of \(h\) and \(r\), and obtain \(\theta(h, r)\) and \(R(h, r)\).
\item Finally, through substitution, we derived
\[\Gh(h, r) = \Gh(\theta(h, r), R(h, r))\,, \]\
and 
\[\Gr(h, r) = \Gr(\theta(h, r), R(h, r))\,.\]
\end{enumerate}
The corresponding contour plots are presented in  Fig.~\ref{fig:m-sigma-R-theta}.

\section{Evaluation of $\tilde \Gh'(0)$ and $\tilde \Gr(0)$.}
\label{sec:s-1st-der}

In this Appendix, we utilize the scaling functions $\bar \Gh(\theta)$ and $\bar \Gr(\theta)$ given by Eqs.~(\ref{eq:m_h_tils-1}) and (\ref{eq:sigma-theta}), respectively, to compute those variables in the scaling form with another argument, $z=h/|r|^{\beta\delta}$, utilized in Eq.~(\ref{eq:m-sigma-sca}).

Before we delve into the specific derivation and explanation of the scaling function of \( z \), there is a key general remark that needs to be highlighted: This function cannot be expressed as a global function but rather as a piecewise function, dependent on the sign of \( r \), specifically \( \tilde{\Gh}_{\text{sgn}(r)}(z) \) and \( \tilde{\Gr}_{\text{sgn}(r)}(z) \). In the main content of this paper, our focus has primarily been on the scenario where \( r \leq 0 \), the region where the coexistence line emerges. As a result, the negative sign has been conventionally omitted, representing \( \tilde{\Gh}(z) \) as \( \tilde{\Gh}_{-}(z) \). However, in this Appendix, we restore this subscript and extend our discussion to encompass the entire \( z \) plane, including the $r>0$ region, for a more comprehensive understanding.

We begin by converting the parametric form from $(R,\theta)$ in Appendix \ref{sec:para} to $(r,\theta)$. To do so, we eliminate $R$ from the expression for $h$ and $\Gh$ provided by Eqs.~(\ref{eq:h}) and (\ref{eq:m}), respectively. This is accomplished using the parametrization of the $r$ variable as stated in (\ref{eq:r}). Consequently, we get
\begin{align}
\label{eq:m2}
\Gh &= |r|^\beta |1-\theta^2|^{-\beta} \bar \Gh(\theta) \equiv |r|^\beta \bar \Gh_2 (\theta) \,, \\
\label{eq:h2}
h &= |r|^{\beta\delta} |1-\theta^2|^{-\beta\delta} \bar h(\theta) \equiv |r|^{\beta\delta} \bar h_2(\theta) \,.
\end{align}
It is imperative to retain the condition $R\geq0$ which acts as constraints on $\theta$ and $r$:
\begin{align}
\label{eq:r-theta-cond}
\left\{
\begin{array}{ll}
1\leq |\theta|<\theta_{\rm max} & (r\leq 0) \\
0<|\theta|<1 & (r>0)
\end{array}
\right. \,.
\end{align}
\begin{figure}[t]
\centering
\includegraphics[bb=0 0 580 370, width=8cm]{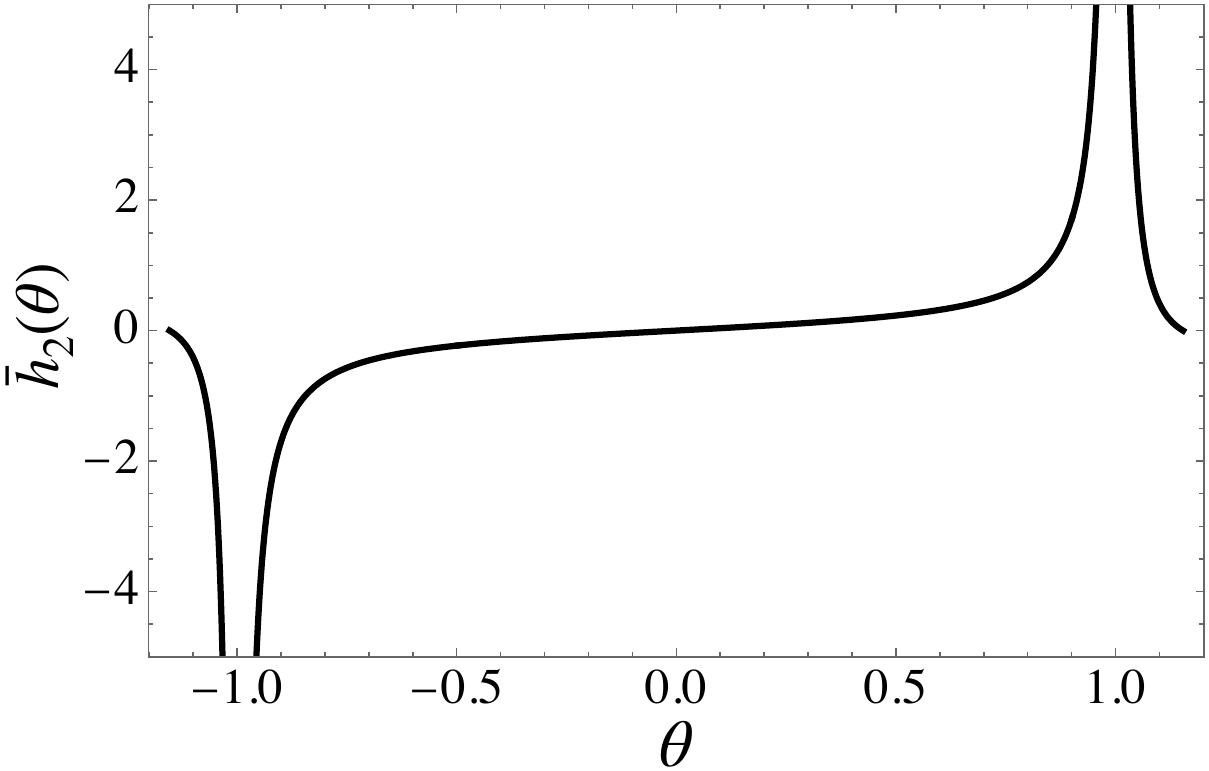}
\caption{A plot of $\bar h_2 (\theta)$}
\label{fig:h2}
\end{figure}
As shown in Fig.~\ref{fig:h2}, $\bar h_2(\theta)$ intersects an arbitrary horizontal line $z$ at multiple points, suggesting the existence of multiple solutions. To identify a unique solution for $\theta$, we apply the condition given in Eq.~(\ref{eq:r-theta-cond}). We introduce a specific function for each $r$ condition $\bar h_{2,{\rm sgn}(r)}^{-1}$:
\begin{align}
\label{eq:theta-z-unique}
\theta &= \bar h_{2,{\rm sgn}(r)}^{-1}(z)
\equiv \bar h_{2}^{-1}(z) \text{,\quad s.t., \ Eq.~(\ref{eq:r-theta-cond})} \,.
\end{align}
Here, the solution of the inverse function $\bar h_{2}^{-1}(z)$ restricted to (\ref{eq:r-theta-cond}) depends on the sign of $r$. If $r\leq0$, then $\bar h_{2,-}^{-1}$ chose the solution of $\bar h_{2}^{-1}(z)$ satisfies $1\leq |\theta|<\theta_{\rm max}$; if $r>0$, then $\bar h_{2,+}^{-1}$ chose $0<|\theta|<1$.
Using (\ref{eq:theta-z-unique}) on the $r\leq 0$ branch, we derive
\begin{align}
\label{eq:theta-max-z}
\pm \theta_{\rm max} &= \bar h_{2,-}^{-1}(z=\pm0) \,,
\end{align}
which will be used to calculate $\Gh$ and $\Gr$ along the first-order phase transition boundary ($r\leq 0$). 

Let us determine the scaling function. Due to the patch structure of the inverse function $\bar h^{-1}_{2,{\rm sgn}(r)}$ as described in (\ref{eq:theta-z-unique}), the scaling functions are defined piecewise based on the sign of $r$.

For magnetization, we express
\begin{subequations}
\begin{align}
\label{eq:m-sca-def-z}
\Gh & \equiv |r|^\beta \tilde \Gh_{{\rm sgn}(r)}(z) \,, \\
\label{eq:m-sca-def-z-2}
\tilde \Gh_{{\rm sgn} (r)}(z) &= \bar \Gh_2 \left( \bar h_{2,{\rm sgn} (r)}^{-1}(z) \right) \,,
\end{align}
\end{subequations}
which extends the scaling form for the negative $r$ region, (\ref{eq:shat-1st}). Note $\tilde \Gh (z)$ employed in the main part is equivalent to the minus branch of Eq.~(\ref{eq:m-sca-def-z-2}). From the scaling function (\ref{eq:m-sca-def-z-2}), we immediately have
\begin{align}
\label{eq:m-z0}
\tilde \Gh_{-}(\pm 0)& = \bar \Gh_2 (\pm \theta_{\rm max}) = \pm 1 \,, 
\end{align}
which is referenced in the main part [below Eq.~(\ref{eq:m-sigma-sca})]. To derive Eq.~(\ref{eq:m-z0}), we utilize Eq.~(\ref{eq:theta-max-z}) to convert specific $z$ value to its corresponding $\theta$ value. Furthermore, we have used the normalization condition for $\Gh_0$ from Eq.~(\ref{eq:norm-1}).

We can calculate the derivative for each region defined by 
 ${\rm sgn}(r)=\pm$:
\begin{align}
\label{eq:sca-m-der}
\tilde \Gh_{{\rm sgn} (r)}'(z) & = \left. \frac{\bar \Gh_2' (\theta) }{\bar h_{2,{{\rm sgn} (r)}}'(\theta)} \right|_{\theta = \bar h_{2,{{\rm sgn} (r)}}^{-1}(z)} \notag\\
&= \left. |1-\theta^2|^{-\gamma} \bar \Gh_h(\theta) \right|_{\theta = \bar h_{2,{{\rm sgn} (r)}}^{-1}(z)}\,,
\end{align}
where $\bar \Gh_h(\theta)$ is defined by $\Gh_h=R^{-\gamma}\bar \Gh_h(\theta)$ with Eq.~(\ref{eq:sus-ch1}). For the middle expression, we have used the formula for the derivative of an inverse function, ${\rm d}f^{-1}/{\rm d}y = 1/f'(x)$, where $y=f(x)$.

Using the explicit form of the derivative given in Eq.~(\ref{eq:sca-m-der}), we compute a specific value used in the main part and (\ref{eq:para-comp}) (note we have dropped off the minus label):
\begin{align}
\label{eq:m-sca-slope-z0}
\tilde \Gh_{-}'(\pm0) & = |1-\theta_{\rm max}^2|^{\gamma} \bar \Gh_h(\pm \theta_{\rm max}) \approx 0.35 \,.
\end{align}

We can apply the same method to the Ising energy density $\Gr$. Eliminating $R$ from the scaling form in terms of $(\theta,R)$ variable, as given by (\ref{eq:sigma-expr}), we have 
\begin{align}
\Gr &= |r|^{1-\alpha} |1-\theta^2|^{\alpha-1} \bar \Gr(\theta) \equiv |r|^{1-\alpha} \bar \Gr_2 (\theta) \,.
\end{align}
This can be further expressed as
\begin{subequations}
\begin{align}
\label{eq:sigma-sca-def-z}
\Gr & \equiv |r|^{1-\alpha} \tilde \Gr_{{\rm sgn}(r)}(z) \,, \\
\label{eq:sigma-sca-def-z-2}
\tilde \Gr_{{\rm sgn}(r)}(z) &= \bar \Gr_2 \left( \bar h_{2,{\rm sgn}(r)}^{-1}(z) \right) \,.
\end{align}
\end{subequations}
Note that the scaling form $\tilde \Gr_{{\rm sgn}(r)}(z)$ is applicable across the entire $r$ region. Evaluating it, we obtain
\begin{align}
\label{eq:sigma_m_z=0}
\tilde \Gr_-(\pm 0) &
= |1-\theta_{\rm max}^2|^{\alpha-1} \bar \Gr(\theta_{\rm max}) \approx -0.66\,,
\end{align}
which completes the derivation of Eq.~(\ref{eq:para-comp}).

The derivative can also be computed:
\begin{align}
\tilde \Gr_{{\rm sgn}(r)}'(z) & = 
\left. \frac{\bar \Gr_2' (\theta) }{\bar h_{2,{\rm sgn}(r)}'(\theta)}  \right|_{\theta = \bar h_{2,{{\rm sgn} (r)}}^{-1}(z)} \notag\\
\label{eq:eps-til-z-der}
& = \left.|1-\theta^2|^{1-\beta} \bar \Gr_h(\theta)  \right|_{\theta = \bar h_{2,{{\rm sgn} (r)}}^{-1}(z)} \,.
\end{align}
Using this expression along with Eqs.~(\ref{eq:m-sca-def-z-2}) and (\ref{eq:sca-m-der}), the Maxwell relation (\ref{eq:maxwell}) can be verified by evaluating both hand sides:
\begin{align}
{\rm sgn}(r) \beta \left[ \tilde \Gh_{{\rm sgn}(r)}(z)-z \delta \tilde{m}_{{\rm sgn}(r)}'(z) \right] = \tilde{\Gr}_{{\rm sgn}(r)}'(z) \,.
\end{align}
This implies
\begin{align}
\tilde{\Gr}_{-}'(\pm 0) = - \beta \tilde \Gh_{-}(\pm 0) = \mp \beta \,,
\end{align}
which has been used under Eq.~(\ref{eq:maxwell}).

\begin{figure*}[t]
\centering
\includegraphics[bb=0 0 1180 760, width=14cm]{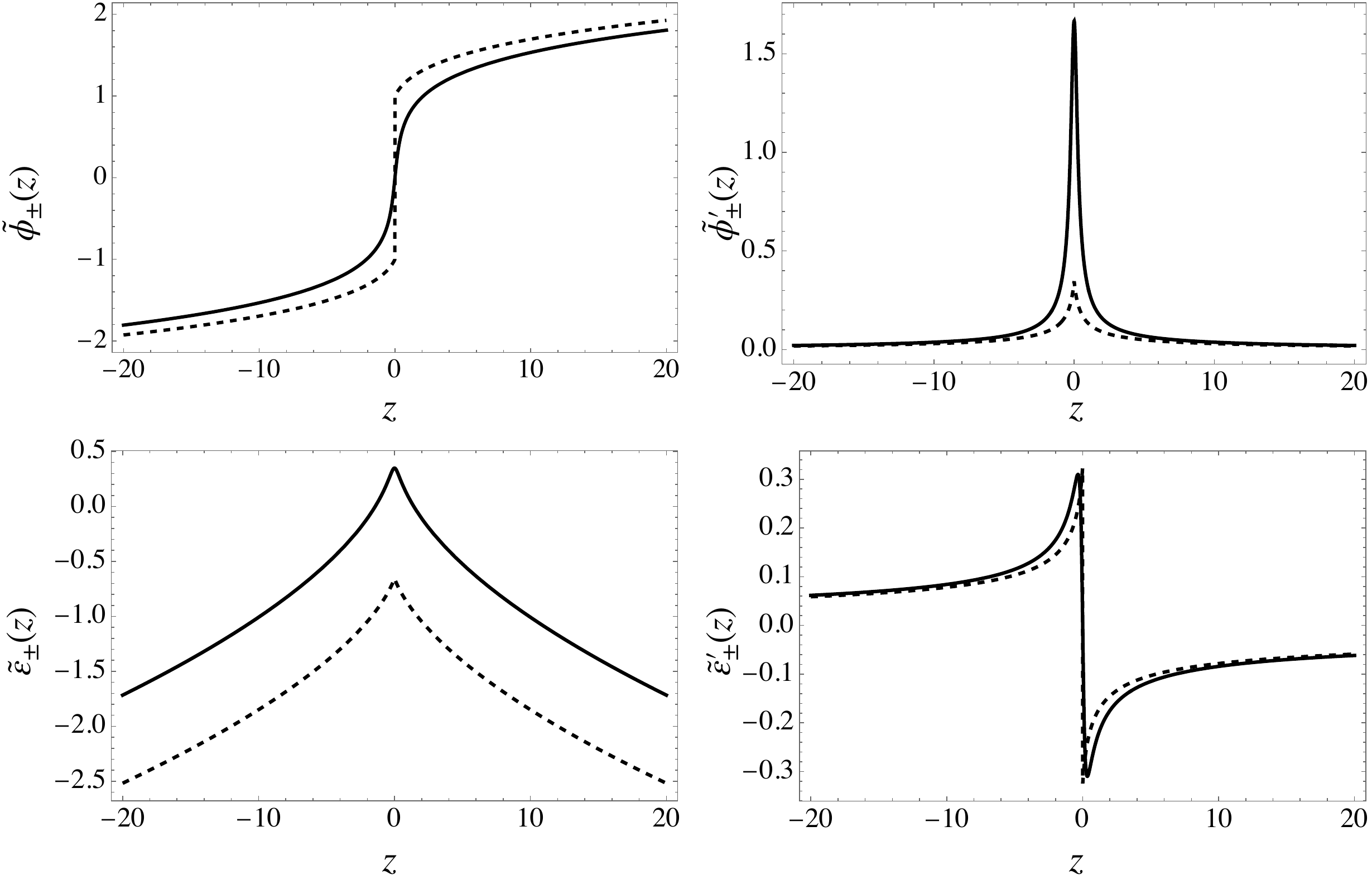}
\caption{Scaling functions $\tilde \Gh_{{\rm sgn}(r)}(z)$ and $\tilde \Gr_{{\rm sgn}(r)}(z)$ along with their derivatives. The solid black line corresponds to ${\rm sgn}(r)=+$ while the dashed line represents ${\rm sgn}(r)=-$.}
\label{fig:scaling-z}
\end{figure*}

In Fig.~\ref{fig:scaling-z}, we display the scaling functions $\tilde \Gh_{\pm}(z)$, $\tilde \Gr_{\pm}(z)$ along with their derivatives $\tilde \Gh_{\pm}'(z)$, $\tilde \Gr_{\pm}'(z)$. 

\section{Topography}
\label{sec:topo}
\begin{figure}[t]
\centering
\includegraphics[bb=0 0 630 340, width=8cm]{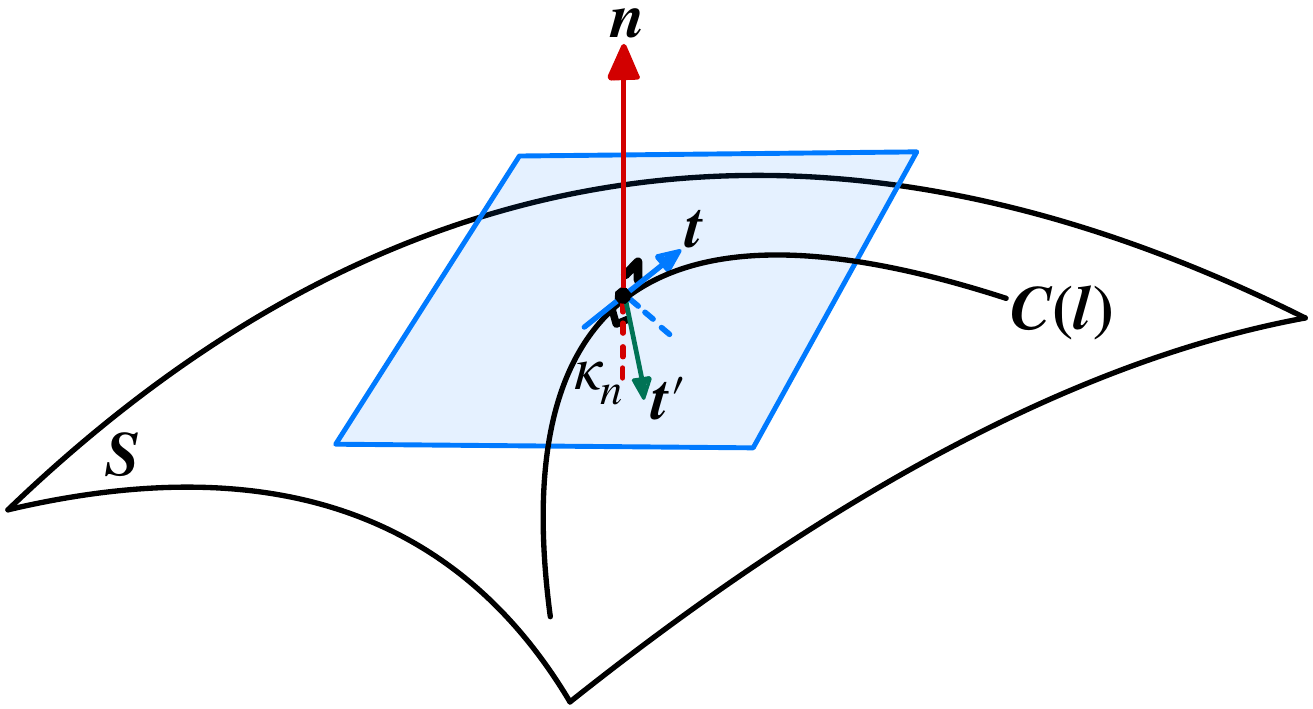}
\caption{Geometric description of the normal curvature $\kappa_n$ (red dashed line) with a normal vector $\nv$ (red arrow), unit tangent vector $\tv$ (blue arrow), and the principal normal vector $\tv'$ (green arrow). Note that $\tv'$ is out of the normal surface (blue shaded area). The dashed lines correspond to the projection of $\tv'$ to the normal vector (red) and the normal surface (blue), respectively.}
\label{fig:normal-curvature}
\end{figure}
We briefly review mathematical tools for exploring the topographic structure of $\hat{s}$ on a general two-dimensional coordinate, $(u,v)=(\mu,\,T)$ or $(r,\,h)$ (see, e.g., Ref.~\cite{DBLP:books/daglib/0019959} for detailed information).

The three-dimensional vector $\Sv=(u,v,\hat s(u,v))$ corresponds to arbitrary points on the curved surface with a height function $\hat s(u,v)$. Let us consider an arbitrary parametric curve embedded on this surface, ${\bm C} (l) = \Sv(u(l),v(l), \hat s (u(l),v(l)))$ (as depicted in Fig.~\ref{fig:normal-curvature}). The unit tangent vector $\tv$ along ${\bm C} (l)$ (represented by the blue arrow) is given by
\begin{align}
\tv = \Sv' \equiv \frac{{\rm d} \Sv}{{\rm d} l} = \Sv_u u' + \Sv_u v'\,,
\end{align}
where we define the line parameter $l$ with a certain normalization, $|\tv^2| = 1$. The tangent plane (blue shaded area) is spanned by $\Sv_u=(1,0,\hat s_u)$ and $\Sv_v=(0,1,\hat s_v)$. The unit surface normal vector (red arrow) can be expressed as
\begin{align}
{\bm n} = \frac{\Sv_u \times \Sv_v}{|\Sv_u \times \Sv_v|} = \frac{(-\hat s_u , -\hat s_v,1)}{\sqrt{1+ \hat s_u ^2 + \hat s_v^2}}\,.
\end{align}
We can also define a normal vector perpendicular to $\tv$, $\tv'$ (green arrow) associated with the curve ${\bm C} (l) $ at the same point that we have defined $\nv$. It is important to note that
$\tv \cdot \tv'=0$ and $\tv \cdot \nv=0$ but $\tv' \cdot \nv\neq 0$ in general (the red dashed line denotes the projection of $\tv'$ to $\nv$).

To determine a tangent vector for ridgelines, we calculate the curvature of the surface $\hat s$ along an arbitrary curve. A curvature along the curve has been given by $\tv'$ so we can take its projection to the normal vector of the surface $\nv$ and know its strength for the surface. This feature is characterized by the normal surface curvature, $\kappa_n \equiv \tv' \cdot \nv$. By using the fact that $\tv \cdot {\bm n}=0$ and $\tv^2 =1$, we can express the normal surface curvature into
\begin{align}
\label{eq:kappa_n_def}
\kappa_{n} = -\Sv' \cdot \nv' = - \frac{{\rm d}\Sv \cdot {\rm d} \nv}{{\rm d} l^2} = \frac{II}{I}\,.
\end{align}
Here, the first and second fundamental forms are defined as
\begin{subequations}
\label{eq:funds}
\begin{align}
I &\equiv {\rm d} \Sv \cdot {\rm d} \Sv = E {\rm d}u^2 + 2F {\rm d}u {\rm d}v + G {\rm d} v^2 \,,\\
II &\equiv - {\rm d} \Sv \cdot \nv = L {\rm d}u^2 + 2M {\rm d}u {\rm d}v + N {\rm d} v^2 \,,
\end{align}
\end{subequations}
where ${\rm d} \Sv = \Sv_u {\rm d}u + \Sv_v {\rm d}v$ and $\Sv_u \cdot \nv =\Sv_v \cdot \nv  =0$. We can compute each component of $I$ and $II$ using an explicit form of the height function $\hat{s}$ given by (\ref{eq:shat-approx}) as follows:
\begin{align}
E= 1 + \hat s_u^2 \,, \quad F= \hat s_u \hat s_v \,, \quad G = 1 + \hat s_v^2 \,.
\end{align}
and 
\begin{align}
(L,M,N)&= (\Sv_{uu},\Sv_{uv},\Sv_{vv}) \cdot \nv = \frac{(\hat s_{uu},\hat s_{uv},\hat s_{vv})}{\sqrt{1+ \hat s_u ^2 + \hat s_v^2}} \,,
\end{align}
where $\Sv_{uu} = (0,0,\hat s_{uu})$, etc.

Substituting the fundamental forms (\ref{eq:funds}) into (\ref{eq:kappa_n_def}), the curvature $\kappa_n$ becomes a function of the slope $\lambda = {\rm d} v / {\rm d}u$ between two infinitesimally separated points $(u, v)$ and $(u+{\rm d} u,v+{\rm d}v)$:
\begin{align}
\label{eq:kappa-n-lambda}
\kappa_n = \frac{L + 2M \lambda + N \lambda^2}{E + 2F \lambda + G\lambda^2} \,.
\end{align}
As the direction of the infinitesimally separated point can be adjusted arbitrarily (due to a $360^\circ$ rotation), the curvature (\ref{eq:kappa-n-lambda}) varies as a function of this direction $\lambda$. The surface's curvature generally has maximal and minimal values in different directions, denoted as $\kappa_{\rm extr.} = \kappa_{\rm min}, \kappa_{\rm max}$ ($\kappa_{\rm min} \leq \kappa_{\rm max}$), which are referred to as the principal curvatures. In what follows, we will examine the general case, except for $\kappa_{\rm min} = \kappa_{\rm max}$. By evaluating ${\rm d}\kappa_n / {\rm d} \lambda = 0$, the expression for the normal curvature (\ref{eq:kappa-n-lambda})  reduces to:
\begin{align}
\label{eq:kappa-ext}
\kappa_{\rm extr.}  = \frac{M+N \lambda}{F+G\lambda} = \frac{L+M \lambda}{E+F\lambda} \,,
\end{align}
from which we can determine the extremum values. By eliminating $\lambda$ from these equations, we arrive at the quadratic equation:
\begin{align}
\label{eq:kappa-ext-2}
\kappa_{\rm extr.}^2 - 2H \kappa_{\rm extr.} +K &=0\,,
\end{align}
where $H=(\kappa_{\rm max}+\kappa_{\rm min})/2$ and $K=\kappa_{\rm max}\kappa_{\rm min}$ are referred to as the mean and Gaussian curvatures, respectively:
\begin{align}
K=\frac{LN-M^2}{EG-F^2} \,,\quad H =\frac{EN+GL-2FM}{2(EG-F^2)}\,.
\end{align}
By solving the quadratic equation (\ref{eq:kappa-ext-2}), we obtain the maximum and minimum curvatures:
\begin{align}
\label{eq:kappa-min-max}
\kappa_{\rm max/min} = H\pm\sqrt{H^2-K} \quad (\kappa_{\rm min} \leq \kappa_{\rm max})\,.
\end{align}
The directions $\lambda$ corresponding to these curvatures refer to the principal directions and can be written in a vector form as $\pv_{\rm max/min}=\pv(\kappa_{\rm max/min})$ with $\pv(\kappa_n)$ given by 
\begin{align}
\label{eq:p_kapa_n}
\pv(\kappa_n) \equiv \left(
\begin{array}{cc}
 u \\
 v 
\end{array}
\right) = \left(
\begin{array}{cc}
 N-\kappa_n G \\
 -(M-\kappa_n F)
\end{array}
\right), \left(
\begin{array}{cc}
 M-\kappa_n F \\
 -(L-\kappa_n E)
\end{array}
\right) \,,
\end{align}
where $\kappa_n$ takes $\kappa_{\rm max}$ or $\kappa_{\rm min}$.

Ridgelines are the lines that an observer who walks along sees the low levels on the left and right sides. The direction across a ridgeline must simultaneously exhibit an extremum and have negative curvature. This local maximum should occur in either of the directions where the curvature is extremized, $\pv_{\rm max/min}=\pv(\kappa_{\rm max/min})$. Consequently, ridge lines (which are not flat along them) are defined as follows \cite{doi:10.1177/027836498300200105}: 
\begin{align}
\label{eq:ridge-line-def}
g_{\rm max/min} \equiv \na \hat s \cdot \pv_{\rm max/min}=0\,, \quad \kappa_{\rm max/min}<0\,,
\end{align}
where $\na=(\partial_u,\partial_v)$. The line represents a valley if the same equality holds for $\kappa_{\rm max/min}>0$.  
A geometric approach to identifying ridgelines can be deduced from Eq.~(\ref{eq:ridge-line-def}): it involves a curve in which the surface's gradient direction is perpendicular to the principle directions.

\section{Specific heat}
\label{sec:specific-heat}
In this Appendix, we elucidate the relationship between specific heat at constant volume and the temperature derivative of specific entropy along $h = \pm 0$,
\begin{align}
\label{eq:dshatdThpm0-expr}
&T_\crit \left( \frac{\partial \hat s}{\partial T} \right)_{h=\pm0} \notag\\
&= \mp B_\Gh  \beta \left( \frac{-\Delta T}{T_\crit} \right)^{\beta-1}
+ B_\Gr (\alpha-1)\left( \frac{-\Delta T}{T_\crit} \right)^{-\alpha} \,,
\end{align}
which is a derivative of the key expression of $\cx{\hat s}$ as introduced in Eq.~(\ref{eq:sTc-T}) or its detailed form Eq.~(\ref{eq:shat-1st-T}). 

The first and the second terms of (\ref{eq:dshatdThpm0-expr}) correspond to the leading and subleading fluctuations generated in the response to the variance of the $r$ variable, the cross susceptibility $\Gh_r$ and the specific heat (at constant $h$) $\Gr_r$, given by (\ref{eq:suscept-cross-1}) and (\ref{eq:suscept-2}), respectively. Since the $h$ variable is fixed, the channel with the strongest criticality in Eq.~(\ref{eq:suscept-h0}), $\Gh_h$ is absent.

\subsection{Preliminary: $\dis \left( \frac{\partial \hat s}{\partial T} \right)_{n}$ and $\dis \left( \frac{\partial \hat s}{\partial T} \right)_{h}$}

\label{sec:specific-heat-pre}
The relationship between the $T$ derivative of $\hat s$ at fixed $h$ and $n$ can be understood based on
\begin{align}
\label{eq:dshatdTn-dshatdTh}
\left( \frac{\partial \hat s}{\partial T} \right)_{n} = \left( \frac{\partial \hat s}{\partial T} \right)_{h} -  \left( \frac{\partial \hat s}{\partial n} \right)_T \left( \frac{\partial n}{\partial T} \right)_h\,.
\end{align}
Using Eq.~(\ref{eq:shat-approx}), we compute the leading (and subleading) order terms of each factor as follows:
\begin{subequations}
\begin{align}
\label{eq:dshatdTh-LO}
\left( \frac{\partial \hat s}{\partial T} \right)_{h} &= \shatGhc \Gh_r \left( \frac{\partial r}{\partial T} \right)_{h} + \shatGrc \Gr_r \left( \frac{\partial r}{\partial T} \right)_{h} \,, \\
\label{eq:dndTh-LO}
\left( \frac{\partial n}{\partial T} \right)_{h} &= A h_\mu \Gh_r \left( \frac{\partial r}{\partial T} \right)_{h} + A r_\mu \Gr_r \left( \frac{\partial r}{\partial T} \right)_{h} \,,\\
\label{eq:dshatdnT-LO}
\left( \frac{\partial \hat s}{\partial n} \right)_{T} &= \frac{\shatGhc}{\left( \frac{\partial n}{\partial \Gh} \right)_{\Gr}} + \cdots = \frac{\shatGhc}{Ah_\mu} + \cdots \,.
\end{align}
\end{subequations}
These expressions show that the contributions from the cross susceptibility $\Gh_r$ on the first term and the second terms of (\ref{eq:dshatdTn-dshatdTh}) cancel each other. Consequently, the leading term in the specific heat at constant volume behaves as the subdominant term along the $r$-axis fluctuation, $c_V  = Tn (\partial \hat s/\partial T)_n \sim (-r)^{-\alpha}$, i.e., the subleading term on Eq.~(\ref{eq:dshatdThpm0-expr}).

In the subsequent sections, we derive a detailed expression for $c_{V}$ in Appendix~\ref{sec:specific-heat-n}. Additionally, in Appendix~\ref{sec:specific-heat-P}, we calculate the specific heat at constant $P$, $c_P = Tn (\partial \hat s/\partial T)_P$, which exhibits different leading behavior $c_P \sim (-r)^{-\gamma}$, compared to those of $c_V$ and Eq.~(\ref{eq:dshatdThpm0-expr}).

\subsection{Specific heat at constant volume}

\label{sec:specific-heat-n}
To complete the leading order expression of $c_V$, we focus on computing (\ref{eq:dshatdnT-LO}), incorporating the next-to-leading order term. To do this, we rewrite it into 
\begin{align}
\label{eq:dsdnT-expr-2}
\left( \frac{\partial \hat s}{\partial n} \right)_{T} & = \frac{\hat s_\mu}{n_\mu} = \frac{\shatGhc (\Gh_h h_\mu + \Gh_r r_\mu) + \shatGrc \Gr_h h_\mu}{Ah_\mu (\Gh_h h_\mu + \Gh_r r_\mu) + Ar_\mu \Gr_h h_\mu} \,,
\end{align}
where higher-order terms proportional to $\propto \Gr_r$ have already been neglected. Keeping in mind that $\Gh_r = \Gr_h$, we can expand the right-hand side of Eq.~(\ref{eq:dsdnT-expr-2}) in terms of $\Gh_r/\Gh_h$:
\begin{align}
\label{eq:dsdnT-NLO}
\left( \frac{\partial \hat s}{\partial n} \right)_{T} & \simeq \frac{\shatGhc}{A h_\mu} + \frac{h_\mu \shatGrc - r_\mu \shatGhc}{A h_\mu^2} \frac{\Gh_r}{\Gh_h} + \cdots \,.
\end{align}

The rest of the analysis parallels Appendix~\ref{sec:specific-heat-pre}. Substituting Eq.~(\ref{eq:dsdnT-NLO}) into the expression of $(\partial \hat s /\partial T)_n$ given by Eq.~(\ref{eq:dshatdTn-dshatdTh}) with Eqs.~(\ref{eq:dshatdTh-LO}) and (\ref{eq:dndTh-LO}), we obtain 
\begin{align}
\label{eq:dsdT-n-general}
\left( \frac{\partial \hat s}{\partial T} \right)_{n} & = \frac{A}{T_{\crit \parallel}^2 n_\crit^2} \left( \frac{\partial r}{\partial T} \right)_{h}^2 \frac{\Gh_h \Gr_r - \Gh_r^2}{\Gh_h} + \cdots \,,
\end{align}
where $(\Gh_h \Gr_r - \Gh_r^2)/\Gh_h \sim (-r)^{-\alpha}$ and we have used 
\begin{align}
\left( \frac{\partial r}{\partial T} \right)_{h} = \frac{h_\mu r_T - h_T r_\mu}{h_\mu} = \frac{1} {\Delta T_1} \,,
\end{align}
with $\Delta T_1$ given below Eq.~(\ref{eq:Bs}). 

Let us apply Eq.~(\ref{eq:dsdT-n-general}) along the coexistence line by setting $h=\pm 0$ with the susceptibilities given by Eq.~(\ref{eq:suscept-cofs}). Specifically, it describes the slope of the curve with a constant $n$ on the $(T,\hat s)$ plane at the point where it intersects with the coexistence line. It becomes clear from the analysis in Appendix~\ref{sec:specific-heat-pre} that the slopes of curves with constant $n$ and at $h=\pm 0$ may differ at the same point, except at the critical point where both become vertical.

The specific heat at constant volume will be 
\begin{align}
\label{eq:cvh0expr}
\frac{\cx{c_V}}{T_\crit^3}  = \frac{K}{\rho^2 w^2 \sin^2 \alpha_1} (-r)^{-\alpha} \,,
\end{align}
which is independent of the sides $h=\pm 0$. The universal factor:
\begin{align}
K = \frac{(\alpha -1) \tilde \Gr(0) \tilde \Gh'(0) - \beta ^2 }{\tilde \Gh'(0)} \,,
\end{align}
is determined by the critical exponents and the values of the scaling function, Eq.~(\ref{eq:para-comp}), to be $K \approx 0.28 $.

\subsection{At constant pressure}
\label{sec:specific-heat-P}

We shall extend the analysis of specific heat by considering a different fixed variable, $c_P=Tn (\partial \hat s/\partial T)_P$. We begin with 
\begin{align}
\label{eq:cp-calc-1}
\left( \frac{\partial \hat s}{\partial T} \right)_P =  \shatGhc \left( \frac{\partial \Gh}{\partial T} \right)_P + \shatGrc \left( \frac{\partial \Gr}{\partial T} \right)_P \,.
\end{align}
Considering $\Gh=\Gh(\mu,\,T)$, we can write 
\begin{align}
\left( \frac{\partial \Gh}{\partial T} \right)_P = \Gh_T + \Gh_\mu \left( \frac{\partial \mu}{\partial T} \right)_P = \Gh_T - \Gh_\mu \hat s\,,
\end{align}
where we have used ${\rm d} P = s {\rm d} T + n {\rm d} \mu$. Substituting this expression and a similar one for $\Gr$ into the starting point, we arrive at a bilinear form, 
\begin{align}
c_P 
& \simeq \frac{n_\crit^2}{T_\crit^3} 
\left(\shatGhc,\shatGrc \right)
\left(
\begin{array}{cc}
\Gh_h & \Gh_r \\
\Gr_h & \Gr_r
\end{array}
\right)
\left(
\begin{array}{c}
 \shatGhc \\
 \shatGrc 
\end{array}
\right)\,,
\end{align}
whose positivity is obvious from the thermodynamic stability. The leading criticality emerges from the (1,1) component, $\propto \Gh_h$. Therefore, the specific heat is approximated as  
\begin{align}
c_{P} & \sim \frac{n_\crit^2}{T_\crit^3}\shatGhc^2 \Gh_h \,,
\end{align}
where $\Gh_h \sim (-r)^{-\gamma}$. 

Along the first-order boundary, we obtain
\begin{align}
\frac{\cx{c_{P}}}{T_\crit^3} 
& =  \frac{(\cos \alpha_1 - \hat s_\crit \sin \alpha_1)^2}{w^2 \sin^2 \alpha_{12}} \tilde \Gh'(0) (-r)^{-\gamma} \,, 
\end{align}
which is also positive since $\tilde \Gh'(0)>0$ checked by Eq.~(\ref{eq:para-comp}).

\section{Specific entropy maximum on the $(n,\,s)$ plane}
\label{sec:n-s_plane}

\begin{figure}[t]
\centering
\includegraphics[bb=0 0 650 850, width=8cm]{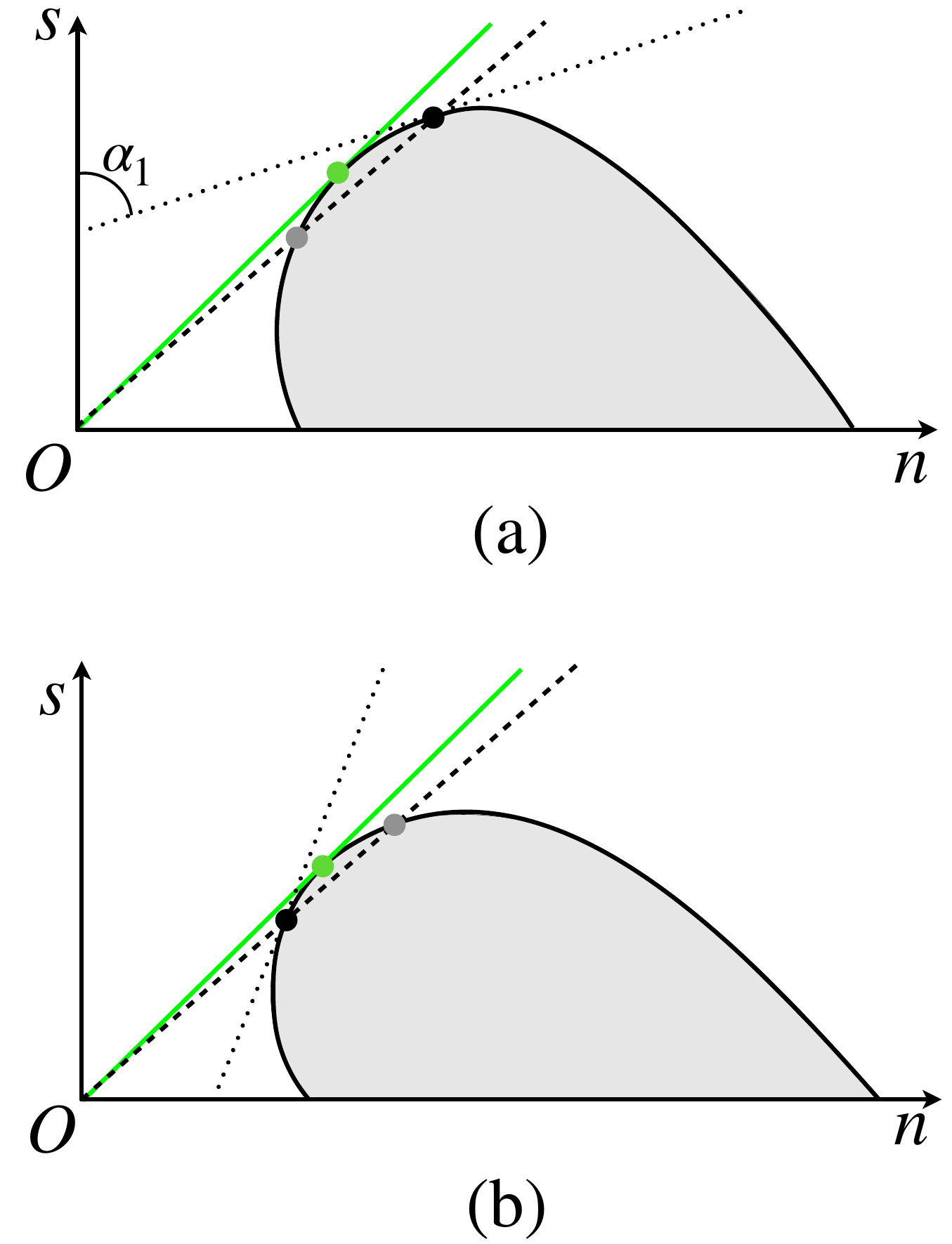}
\caption{Two possible scenarios for the location of the maximum of the specific entropy (green dot) with respect to the critical point (black dot): (a) $\hat s_{\rm c}>\cot \alpha_1$ and (b) $\hat s_{\rm c}<\cot \alpha_1$ (see text).}
\label{fig:n-s_schematic}
\end{figure}

We can also understand the maximum of the specific entropy on the coexistence region boundary in the $(n,\,s)$ plane. 

An isentrope with a fixed value of $\hat s$ is a straight line passing through the origin.
In Fig.~\ref{fig:n-s_schematic}, we illustrate two isentropes, $s = \hat s_{\rm c} n $ (dashed line) and $s = \hat s_{\rm max} n $ (green line). 

The boundary separating the coexistence region (shaded) from the uniform phases is represented by a black curve on this plane, defined by   $h(n,\,s)=\pm0$, $r<0$ near the critical point.%
\footnote{
The boundary is a parametric plot, $s=s(r)$ and $n=n(r)$, where the right-hand sides are given by Eqs.~(\ref{eq:s-def-expr}) and (\ref{eq:n-def-expr}) with Eq.~(\ref{eq:m-sigma-sca}) at $h=\pm0$.\label{fn:boundary-crit}}
The critical point (black dot) divides the curve into two branches, each serving as the boundary for a different uniform phase. The critical isentrope and the boundary also intersect at another point, the ``critical double" (gray dot), as introduced in Fig.~\ref{fig:nonmono}.

The maximum specific entropy on the coexistence boundary is demonstrated by the existence of an isentrope tangential to the boundary. In Fig.~\ref{fig:n-s_schematic}, this tangent line and its corresponding tangent point are shown as a green line and a green dot, respectively. It is straightforward to compute the maximum specific entropy by identifying the tangent isentrope from the critical EOS given in Sec.~\ref{sec:IsingQCD} (see also footnote~\ref{fn:boundary-crit}). The results agree with what we discussed in Sec.~\ref{sec:s/n-first} on the $(\mu,\,T)$ plane.

Figures \ref{fig:n-s_schematic}(a) and \ref{fig:n-s_schematic}(b) show the two scenarios where the maximum specific entropy can occur. Each illustrates the maximum on a different branch which can also be classified according to the formula (\ref{eq:lambda-hill-T}). On the $(n,\,s)$ plane, $\hat s_{\rm c}$ and $\cot \alpha_1$ have a geometric interpretation: the former represents the slope of the critical isentrope (dashed line), while the latter represents the slope of the tangent line to the boundary at the critical point (dotted line). One can show
\begin{align}
\label{eq:dsdn_hc}
\left. \left(\frac{\partial s}{\partial n}\right)_h \right|_{
\crit} = \cot \alpha_1\,,
\end{align}
in contrast to the vertical or horizontal slopes on a different plane including one intensive variable, e.g., $(T,\, \hat s)$ (see Fig.~\ref{fig:nonmono}) or $(n,\, T)$ diagrams, respectively.

Contour classification becomes more intuitive on the $(n, s)$ plane: straight lines above $s = \hat{s}_{\rm max} n$ do not intersect the boundary (class I); lines between $s = \hat{s}_{\rm max} n$ and $s = \hat{s}_{\crit} n$ intersect only one branch of the boundary (class II); lines below $s = \hat{s}_{\crit} n$ intersect both branches of the boundary (class III).

\bibliographystyle{apsrev4-2}
\bibliography{refs.bib}

\end{document}